\documentclass[twocolumn,prb,aps,superscriptaddress]{revtex4-1}
\usepackage[T1]{fontenc}
\usepackage[latin9]{inputenc}

\usepackage{amsmath}
\usepackage{amssymb}
\usepackage{bbm}
\usepackage{braket}
\usepackage{xcolor}
\usepackage[normalem]{ulem}

\numberwithin{equation}{section}

\allowdisplaybreaks

\usepackage{graphicx}
\usepackage[colorlinks=true]{hyperref}  
\hypersetup{
    bookmarks=true,         
    unicode=false,          
    pdftoolbar=true,        
    pdfmenubar=true,        
    pdffitwindow=false,     
    pdfstartview={FitH},    
    pdftitle={Friedel oscillations and Majorana zero modes in inhomogeneous superconductors},    
    pdfauthor={Lauke, Scheurer, Poenicke, Schmalian},     
    pdfsubject={},   
    pdfcreator={},   
    pdfproducer={}, 
    pdfkeywords={} {} {}, 
    pdfnewwindow=true,      
    colorlinks=true,       
    linkcolor=blue, 
    citecolor=blue,        
    filecolor=magenta,      
    urlcolor=blue           
}

\newcommand{\equref}[1]{Eq.~(\ref{#1})}

\newcommand{\secref}[1]{Sec.~\ref{#1}}

\newcommand{\figref}[1]{Fig.~\ref{#1}}
\newcommand{\refcite}[1]{Ref.~\onlinecite{#1}}
\newcommand{\refscite}[1]{Refs.~\onlinecite{#1}}

\newcommand{\pdagger}{{\phantom{\dagger}}}

\renewcommand{\approx}{\simeq}
\renewcommand{\Re}{\text{Re}}
\renewcommand{\Im}{\text{Im}}
\renewcommand{\vec}[1]{\boldsymbol{#1}}

\begin{document}

\title{Friedel oscillations and Majorana zero modes in inhomogeneous superconductors}
\author{Lars Lauke}
\affiliation{Institute for Theory of Condensed Matter, Karlsruhe Institute of Technology (KIT), 76131 Karlsruhe, Germany}
\affiliation{Institute for Solid State Physics, Karlsruhe Institute of Technology (KIT), 76131 Karlsruhe, Germany}
\author{Mathias S.~Scheurer}
\affiliation{Institute for Theory of Condensed Matter, Karlsruhe Institute of Technology (KIT), 76131 Karlsruhe, Germany}
\affiliation{Department of Physics, Harvard University, Cambridge MA 02138, USA}
\author{Andreas Poenicke}
\affiliation{Institute for Theory of Condensed Matter, Karlsruhe Institute of Technology (KIT), 76131 Karlsruhe, Germany}
\affiliation{Institut f\"ur Theoretische Festk\"orperphysik, Karlsruhe Institute of Technology, D-76131 Karlsruhe, Germany}
\author{J\"org Schmalian}
\affiliation{Institute for Theory of Condensed Matter, Karlsruhe Institute of Technology (KIT), 76131 Karlsruhe, Germany}
\affiliation{Institute for Solid State Physics, Karlsruhe Institute of Technology (KIT), 76131 Karlsruhe, Germany}

\begin{abstract}
We study the modulations of the superconducting order parameter in the vicinity of edges, magnetic and non-magnetic impurities by self-consistently solving the gap equations of a system with competing interactions in the Cooper channel. It is shown that the presence or absence of Friedel-like oscillations of the superconducting order parameter crucially depends on its symmetry and can, hence, be used to obtain information about the symmetry properties of the condensate. Furthermore, the appearance of competing order parameters at inhomogeneities is discussed. We show that this can lead to the presence of a topologically trivial region close to the boundary of a system that is topologically nontrivial in its bulk. The resulting shift in position of the Majorana bound states is demonstrated to significantly affect its signatures in Josephson-junction experiments. We discuss Josephson scanning tunneling microscopy as a probe to resolve the Friedel-like oscillations as well as the spatial texture of competing $s$-wave superconductivity and Majorana bound states in the vicinity of the edge of the system.
\end{abstract}
\maketitle
\section{Introduction}
\label{sec:intro}
While the assumption of perfect lattice-translation symmetry facilitates the theoretical treatment of solid state systems, it is inevitably broken in reality due to the presence of impurities or defects in the crystal and the fact that samples are of finite size. In the context of superconductivity, the study of impurities has a long history \cite{Review1Impurities}. The classic result \cite{AT,ATAG1,ATAG2,BWState,GolubovMazin} that the impact of non-magnetic disorder on the superconducting transition temperature $T_c$ crucially depends on the symmetry of the order parameter already shows that disorder can be seen as a blessing in disguise: It allows to obtain information about the symmetries of the order parameter, although this approach requires care as has become clear in recent years\cite{SpinOrbitLocking,PhysRevB.90.134517,PhysRevB.91.054501,PhysRevB.92.014518}. While these studies assume a homogeneous superconducting order parameter, the self-consistent solution of the gap equation of an $s$-wave superconductor shows Friedel-like oscillations, with wavevector set by the Fermi momentum $k_F$, in the vicinity of a non-magnetic impurity as was first shown in \refcite{FetterOldImpurity}. 
The associated correction to the reduction of the transition temperature, suppressed by a factor $\propto T_c/E_F$ ($E_F$ denotes the Fermi energy), was analyzed in \refscite{Tsuzuki,HeinrichMagnetic,Kuemmel,Schlottmann} for $s$-wave superconductors in the presence of magnetic impurities. While formally of the order $T_c/E_F$, these Friedel oscillations of the superconducting gap can be numerically sizable in low-dimensional systems\cite{tanaka1,tanaka2}. Unconventional even-parity, spin-singlet states were analyzed in \refscite{WalkerOsc,FranzTc}. Similarly to impurities, $s$-wave superconductors are also known to exhibit Friedel-like oscillations at the surfaces of the system \cite{GiamarchiSurfaceSC,TanakaPRB,tanaka1}.

The physics of inhomogeneities due to sample edges has received enormous attention in recent years due to the emergence of topological insulators and superconductors\cite{HasanKaneReview,AliceaReview,Bernevig}: The celebrated ``bulk-boundary correspondence'' relates topological invariants, which are discrete functions of the Hamiltonian describing the bulk of the system, to the presence of zero-energy modes localized at the edge of the system. In the case of topological superconductors, the charge-conjugation symmetry, that naturally emerges in the Bogoliubov-de Gennes (BdG) formalism, leads to a reality constraint on the wave function of the associated zero-energy modes, which are thus referred to as Majorana bound states (MBSs). 
While the bulk-boundary correspondence has been exhaustively analyzed for noninteracting, quadratic, Hamiltonians, it requires further theoretical investigation in the presence of interactions\cite{BudichReview}.   
On the experimental side, the unambiguous detection and controlled manipulation of MBSs is the current central open question which is not only interesting in its own right but also constitutes the first step towards topological quantum computation\cite{AliceaReview,Bernevig} and provides information about the underlying pairing mechanism \cite{ScheurerOxide,ScheurerSymmetryMechanism} of superconductivity. One of the most promising routes towards clear identification of MBSs and, potentially, qubit readout\cite{AliceaReview} is based on the MBS-mediated Josephson effect \cite{Kitaev,TRSMajoranaJosephson}.

In this paper, we show that the Friedel-like oscillations of the superconducting order parameter at the edge of one-dimensional (1D) and two-dimensional (2D) systems and in the vicinity of both magnetic and non-magnetic inhomogeneities in 1D are only present in the even-parity channel but not for an odd-parity order parameter. Consequently, the observation (the absence) of oscillations of the order parameter on the length scale $1/k_F$ indicates that the order parameter is even parity (odd parity). This also allows to gain information about the topology of superconductivity as a gapped system with odd-parity pairing realizes a time-reversal invariant topological superconductor given its Fermi surfaces enclose an odd number of time-reversal invariant momenta and a trivial phase in the case of even-parity pairing\cite{FuBergProposal,Sato}. 
Thus, we present another approach to utilize inhomogeneities in order to access both the symmetry and topology of the superconducting order parameter.
Potential candidate systems for time-reversal invariant topological superconductivity include the doped topological insulator Cu$_x$Be$_2$Se$_3$\cite{FuBergProposal}, (quasi-)1D wires\cite{InteractingWires,dwaveWire,AnotherWirePaper,BuddichWire}, bilayer Rashba systems\cite{PhysRevLett.108.147003}, and oxide heterostructures \cite{ScheurerOxide}. The superfluid analogue\cite{BWState,BWStateMajorana} of this phase is believed to be realized in the B phase of $^3$He.

\begin{figure}[tb]
\begin{center}
\includegraphics[width=0.8\linewidth]{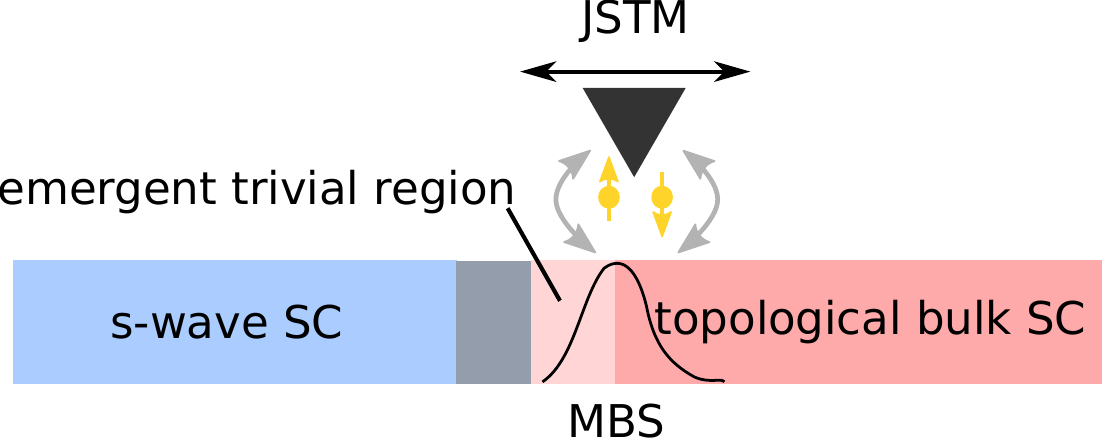}
\caption{Schematic of Josephson junction between an $s$-wave superconductor (blue) and an interaction-induced topological superconductor (red) for the detection of MBSs\cite{Kitaev,TRSMajoranaJosephson}. In addition, we show a JSTM\cite{MartinProp,FirstExperimentJSTM,FabricationOfTips,Hamidian:2016aa,Yazdani,SuperconductingSTM} tip that allows to probe the spatial texture of the order parameter and of the MBS wave function as we discuss in this paper.}
\label{BasicSetup}
\end{center}
\end{figure}

In addition, we provide an example of a modification of the conventional bulk-boundary correspondence due to competing interactions that can significantly influence Josephson tunneling experiments aimed at identifying MBSs.
To this end, we focus on ``intrinsic'' topological superconductors, i.e., systems where the nontrivial topology arises spontaneously from the internal interactions (applies to  Refs.~\onlinecite{InteractingWires,FuBergProposal,ScheurerOxide,BWState,PhysRevLett.108.147003}). For concreteness, we assume the interaction to be dominated by the $p$-wave channel, but  closely followed by the competing $s$-wave instability. By construction, this yields purely $p$-wave, time-reversal invariant topological superconductivity in the bulk; however, the broken inversion symmetry at the edge allows the competing $s$-wave order parameter to emerge locally and leads to a topologically trivial region with width of the order of the coherence length $\xi$ between the topological superconductor and the vacuum. Consequently, the MBSs are not localized at the edge of the system, as expected from the bulk-boundary correspondence, but are ``pushed away'' from it by a length of the order of $\xi$. While this seems like a minor modification at first sight, it can change the signal in the MBS-induced Josephson effect\cite{TRSMajoranaJosephson} between the system and an $s$-wave superconductor, see \figref{BasicSetup}, dramatically: 
Depending on parameters, the current-phase relation can more closely resemble that of a conventional Josephson junction based on two topologically trivial $s$-wave superconductors despite the presence of MBSs and a topologically nontrivial bulk phase.

We finally discuss Josephson scanning tunneling microscopy\cite{MartinProp,FirstExperimentJSTM,FabricationOfTips,Hamidian:2016aa,Yazdani,SuperconductingSTM} (JSTM) as an experimental tool to image the spatial variations of the superconducting order parameter in the vicinity of defects and at the edge, see \figref{BasicSetup}. We generalize the analysis of \refcite{SuperconductingSTM} to finite voltage drops across the junction and also take into account the contribution arising from MBSs. Note that the latter is not captured by the Green's function formalism used in \refcite{SuperconductingSTM} due to the MBS-induced ground state degeneracy. We also analyze the expected JSTM signature of the MBSs that are pushed further into the bulk of the system due to the presence of the competing topologically trivial instability at the edge of the system. We demonstrate that JSTM is a very powerful tool not only to identify the presence of MBSs but also to reveal the spatial profile of the MBS wave function -- even in the presence of competing interactions.

\section{Model and formalism}
\label{ModelIntro}
In this section, we introduce the model we use in this paper to analyze the behavior of superconductivity in the vicinity of inhomogeneities and describe our procedure for solving the corresponding gap equations.

\subsection{Hamiltonian for inhomogeneous superconductors}
In this work, we focus on a single-band lattice Hamiltonian of the form 
\begin{equation}
\hat{\mathcal{H}} = \hat{\mathcal{H}}_{\text{0}} + \hat{\mathcal{H}}_{\text{imp}} + \hat{\mathcal{H}}_{\text{pairing}}\text{.} \label{Hamiltonian}
\end{equation}
The kinetic contributions are represented by 
\begin{equation}
\hat{\mathcal{H}}_{\text{0}} = -t\sum_{\langle ij\rangle}\sum_\sigma c_{i\sigma}^{\dagger} c^\pdagger_{j\sigma} - \mu \sum_{i}\sum_\sigma c_{i\sigma}^{\dagger}c^\pdagger_{i\sigma} \text{,} \label{Hopping}
\end{equation}
where $c_{i\sigma}^{(\dagger)}$ annihilates (creates) an electron of spin \mbox{$\sigma=\{\uparrow,\downarrow\}$} at site $i$. Here, we only consider hopping between nearest-neighbours $\langle ij\rangle$ with hopping amplitude $t$. The boundary conditions of the system will be chosen as \textit{open} (OBC) or \textit{periodic} (PBC), depending on the specific questions at hand. The chemical potential is given by $\mu$. The second term of the Hamiltonian (\ref{Hamiltonian}),
\begin{equation}
\hat{\mathcal{H}}_{\text{imp}} = \sum_{i} \sum_{\sigma\sigma'} c_{i\sigma}^{\dagger} U_{i}^{\sigma\sigma'} c^\pdagger_{i\sigma'}, \label{ImpurityHam}
\end{equation}
allows for magnetic and non-magnetic defects. We take the defects to be localized at a single site $i=i_0$, $U_{i}^{\sigma\sigma'} = u_0 (\hat{\sigma}^{m})_{\sigma\sigma'} \delta_{i,i_0}$ using $u_0$ to denote the strength of the defect. The appropriate spin-structure of the impurities is chosen as $\hat{\sigma}^{m} = \hat{\sigma}^{z}$ and $\hat{\sigma}^{m} = \hat{\sigma}^{0}$ for magnetic and non-magnetic impurities, respectively. The pairing Hamiltonian
\begin{align}
\hat{\mathcal{H}}_{\text{pairing}} = \sum_{\eta=s,t} V^{\eta}& \sum_{i} \biggl(\sum_{\delta\sigma\sigma'} c_{i+\delta,\sigma}^{\dagger}\,\chi_{\sigma'\sigma}^{\eta *} (\delta)c_{i,\sigma'}^{\dagger} \biggr)\nonumber \\
& \times \biggl(\sum_{\tilde{\delta}\tilde{\sigma}\tilde{\sigma}'} c_{i,\tilde{\sigma}'}\,\chi_{\tilde{\sigma}'\tilde{\sigma}}^{\eta} (\tilde{\delta})c_{i+\tilde{\delta},\tilde{\sigma}}\biggr) 
\end{align}
allows for superconducting pairing in the appropriate spin channels corresponding to singlet $s$-wave ($\eta = s$) and triplet $p$-wave ($\eta = t$) with coupling constants $V^{\eta}$. The spin- and spatial-structure leading to \textit{on-site} $s$-wave ($\delta=0$) and \textit{nearest-neighbour} $p$-wave ($\delta=\pm 1$) pairing are encoded in the spin matrices $\chi_{\sigma\sigma'}^{\eta}(\delta)$. The $s$-wave pairing interaction shall be taken to be momentum independent, i.e.~of the form $\chi_{\sigma\sigma'}^{s} = \Delta^{s} (i\hat{\sigma}^{y})_{\sigma\sigma'}$ and, unless otherwise stated, the $p$-wave pairing interaction in momentum space is chosen as $\chi^{t} \propto \vec{d}(\vec{k})\cdot \vec{\sigma}i\hat{\sigma}^{y}$, with triplet vector $\vec{d}(\vec{k}) \propto (-\sin k_y, \sin k_x, 0)^{T}$.

We treat the interacting problem in the mean-field approximation focusing on the Cooper channel and introduce
\begin{equation}\label{eq:Delta}
\Delta_{i}^{\eta} = V^{\eta} \sum_{\delta\sigma\sigma'} \langle c_{i,\sigma'}c_{i+\delta,\sigma}\rangle \chi_{\sigma'\sigma}^{\eta}(\delta)
\end{equation}
as the gap function.
We bring the resulting mean-field Hamiltonian into the standard BdG form\cite{bogoliubov}
\begin{equation}\label{eq:BdG}
\hat{\mathcal{H}}_{\text{MF}} = \frac{1}{2}\sum_{ij}\sum_{\sigma\sigma'} \hat{\Psi}_{i\sigma}^{\dagger} \underbrace{\begin{pmatrix}
\hat{h}_{ij}^{\sigma\sigma'} & \hat{\Delta}_{ij}^{\sigma\sigma'} \\ \bigl(\hat{\Delta}_{ij}^{\sigma\sigma'}\bigr)^{\dagger} & -\left(\hat{h}_{ij}^{\sigma\sigma'}\right)^{T}
\end{pmatrix}}_{\equiv \hat{\mathcal{H}}_{\text{BdG}}} \hat{\Psi}_{j\sigma'} \text{,}
\end{equation}
with BdG Hamiltonian $\hat{\mathcal{H}}_{\text{BdG}}$ and Nambu spinor \mbox{$\hat{\Psi}_{i\sigma} = \bigl( c_{i\sigma}, c_{i\sigma}^{\dagger}\bigr)^T$}. All single-particle contributions, $\hat{\mathcal{H}}_{\text{0}}$ and $\hat{\mathcal{H}}_{\text{imp}}$, are incorporated into the single-particle Hamiltonian $\hat{h}$, while $\hat{\Delta}$ results from $\hat{\mathcal{H}}_{\text{pairing}}$. 
\subsection{Gap equation}
The problem is now reduced to solving the BdG equations
\begin{equation}
\hat{\mathcal{H}}_{\text{BdG}} \, \phi_E = E \, \phi_E \text{,}
\end{equation}
with eigenmodes $\phi_E$. The particle- ($\phi^{(e)}$) and hole-amplitudes ($\phi^{(h)}$) are the Nambu-components of the eigenvectors $\phi_E = (\phi_{E}^{(e)},\phi_{E}^{(h)})^{T}$. The intrinsic redundancy of the BdG Hamiltonian, having effectively doubled the degrees of freedom, leads to the charge-conjugation symmetry
\begin{equation}
\hat{\Xi} \, \hat{\mathcal{H}}_{\text{BdG}}\, \hat{\Xi}^{\dagger} = -\hat{\mathcal{H}}_{\text{BdG}}\text{,}
\end{equation}
where $\hat{\Xi}= \hat{\tau}^{x} C$ is the charge conjugation operator with the first Pauli matrix $\hat{\tau}^{x}$ in particle-hole space and the operator $C$ for complex conjugation. Due to the charge-conjugation symmetry of the BdG equations, a solution $\phi_E$ with energy $E$ is related to a solution $\phi_{-E}$ with energy $-E$ by 
\begin{equation}
\phi_{-E} = \hat{\Xi}\phi_{E}.
\end{equation}
By performing a mode expansion of the spinor
\begin{align}\label{eq:spinorexp}
\hat{\Psi}_{\alpha} =& \begin{pmatrix}
c_{\alpha}\\ c_{\alpha}^{\dagger}
\end{pmatrix} = \sum_{E} (\phi_E)_{\alpha} a_{E}\nonumber \\ =& \sum_{E>0} \biggl((\phi_E)_{\alpha} a_{E}  + \underbrace{(\phi_{-E})_{\alpha} a_{-E}}_{=\left(\hat{\Xi}\phi_E\right)_{\alpha}a_{E}^{\dagger}}\biggr) \text{,}
\end{align}
where $\alpha = \{i,\sigma\}$ is a shorthand for site- and spin-indices, we diagonalize the BdG Hamiltonian. The equivalence of creating a hole and annihilating a particle, i.e. $a_{E} = a_{-E}^{\dagger}$, has been used in \equref{eq:spinorexp} to reduce the expansion to a sum over positive eigenenergies. Using the mode expansion to evaluate the average $\langle c_{\alpha'}c_{\alpha}\rangle$ in \equref{eq:Delta} and introducing an energy cut-off $\epsilon_c$  leads to the inhomogeneous gap equation
\begin{equation}\label{eq:selfcongap}
\Delta_{i}^{\eta} = V^{\eta} \sum_{\delta\sigma\sigma'} \sum_{0<E<\epsilon_c} (\phi_E)_{i\sigma'}^{(e)} (\phi_{E}^{*})_{i+\delta,\sigma}^{(h)} \,\, \chi_{\sigma'\sigma}^{\eta}(\delta) [1 - 2f(E)] \text{.} 
\end{equation}
At $T=0$, which we shall be considering for the remainder of this paper, the Fermi-distribution function \mbox{$f(E) = [e^{\beta E} + 1]^{-1}$} vanishes, due to the lack of thermal excitations at zero temperature.
The results shown below are obtained by solving \equref{eq:selfcongap} via iteration, i.e., iteratively inserting an approximation of $\Delta_{i}^{\eta}$ into the right-hand side of the equation and recalculating its value from \equref{eq:selfcongap} until convergence is achieved. The convergence criterion we use is demanding that the relative change in $\Delta_{i}^{\eta}$ from one iteration to the next one be smaller than $10^{-4}$.

\section{One spatial dimension}
\label{1DResults}
We start our discussion of the numerical solution of the superconducting self-consistency equations (\ref{eq:selfcongap}) by focusing on 1D systems since the oscillatory features in the order parameter are most pronounced and the results can be more conveniently illustrated in one spatial dimension.

\subsection{Competition of singlet and triplet order parameters at edges}
\label{FriedelOscillations}
In the presence of inversion symmetry in the normal state, the superconducting order parameter that assumes finite values below the superconducting transition temperature must be either pure singlet or pure triplet\cite{sigristueda}
However, once inversion is broken in the normal state, even-parity (singlet) and odd-parity (triplet) pairing can and, in general, will mix. 

\begin{figure}[tb]
\includegraphics[width=\linewidth]{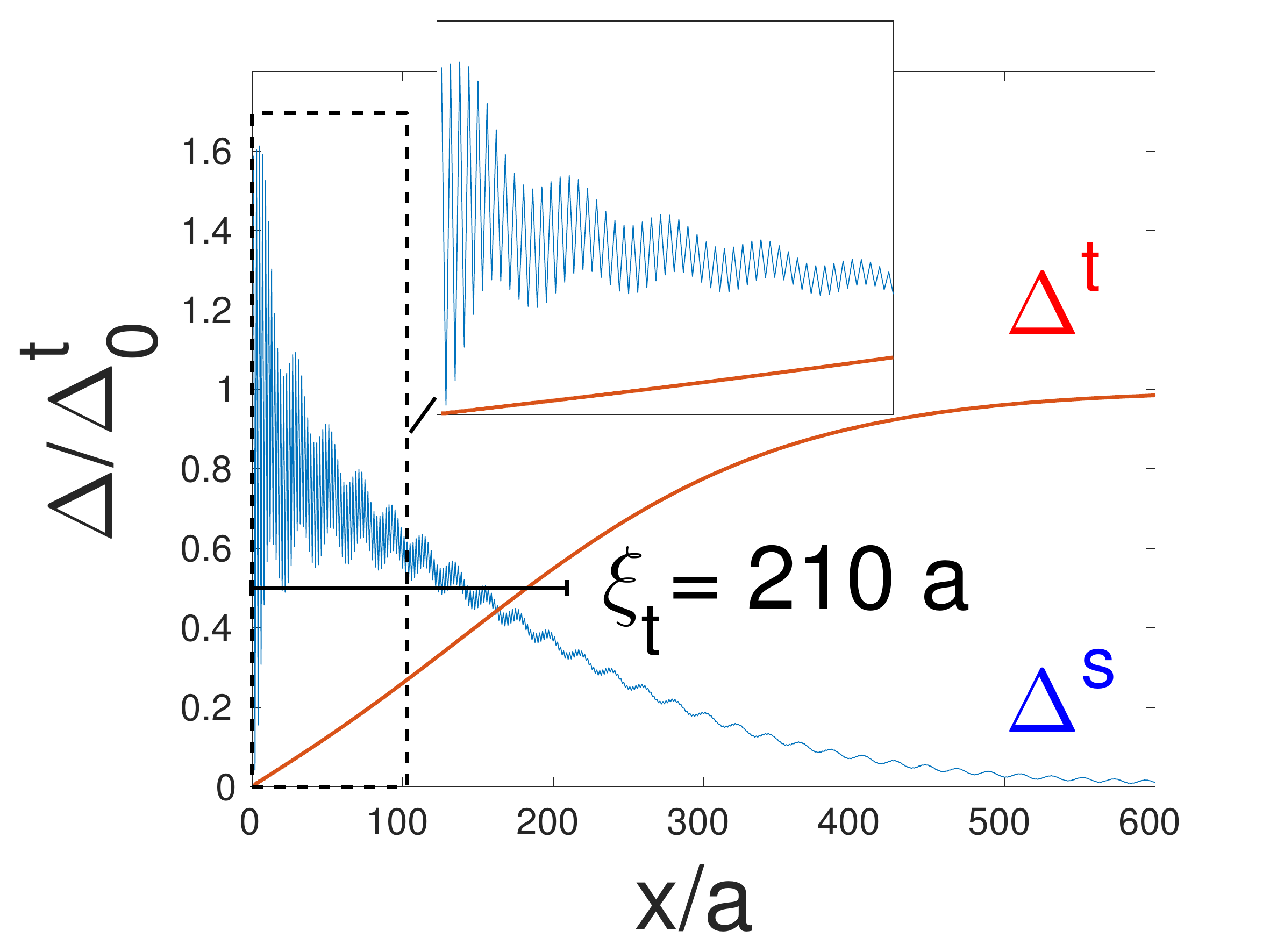}
\caption{Spatial structure of competing $s$-wave (blue) and $p$-wave (red) order parameters at a boundary ($x=0$) for a system of $L=3000$ sites at half-filling ($\mu=0$) for coupling constants $V^{s}=0.36$, $V^{t}=0.38$. The order parameters are normalized to the bulk value $\Delta_{0}^{t}$ of the $p$-wave pairing. The order parameters at the other edge at $x=L$ behave in the same fashion.}
\label{fig:CompetingOP}
\end{figure}

Inversion symmetry can be broken globally by the crystal structure itself or by the environment of a low-dimensional system (e.g., the substrate of a 2D thin film). Alternatively, it can be broken locally at edges and inhomogeneities of the sample. In this work, we will entirely focus on the local breaking of inversion symmetry. As a first example, let us consider a system with dominant triplet pairing in the bulk and allow for a non-vanishing singlet component ($V^t > V^s \neq 0$). The spatial texture of the superconducting order parameter at one of the edges of the system, as found from solving the superconducting gap equations numerically, is shown in \figref{fig:CompetingOP}. The local structure of the $p$-wave order parameter $\Delta^t$ exhibits the expected behavior: a constant bulk value and a decrease towards the boundary on the scale of its coherence length $\xi_t$. Furthermore, a non-vanishing singlet component is present in the vicinity of the boundary, due to the locally broken inversion symmetry explained above. Note, we do not consider the possibility of odd-frequency pairing near the edge\cite{Tanaka}. For the model considered here, with two almost degenerate even frequency order parameters, this seems to be a natural assumption.

The most prominent feature of the local structure of the $s$-wave order parameter are the pronounced \textit{Friedel oscillations}\cite{WalkerOsc} that are entirely absent in the $p$-wave component. These oscillations are not accessible to the quasi-classical approaches, such as the Eilenberger equations\cite{eilenberger}, due to the averaging over energies, leaving the coherence length as the only relevant length scale. The additional shorter oscillation period resolved by our microscopic formalism is set by $k_F$ and is of the order of the lattice spacing $a$. The larger oscillation period, intermediate between $k_F$ and the coherence length $\xi_s$ of the $s$-wave state, is due to the finite energy cut-off $\epsilon_c$ and disappears for larger $\epsilon_c$ or as we move away from half-filling.

\begin{figure}[tb]
\includegraphics[width=\linewidth]{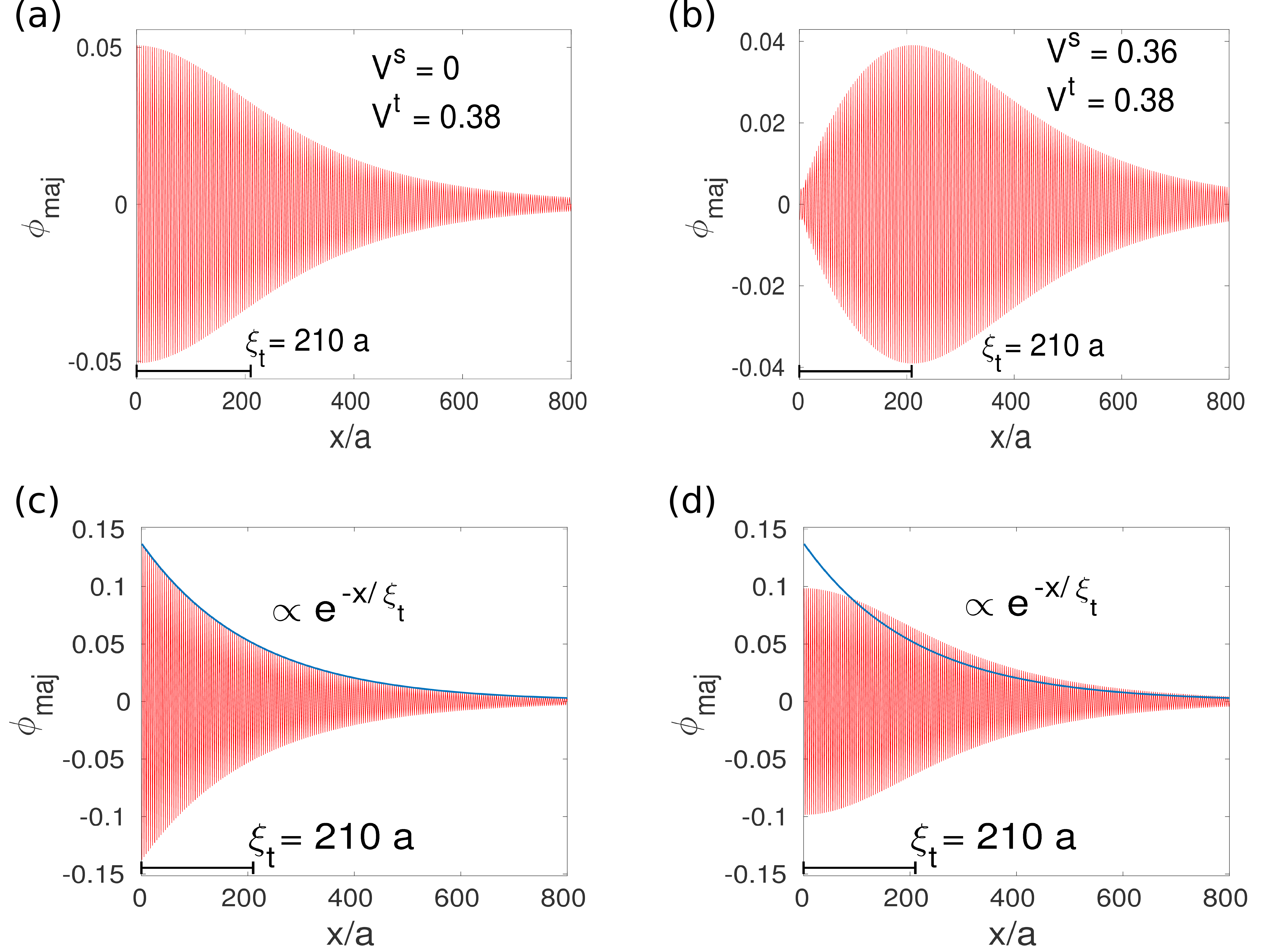}
\caption{MBS wave function at the left edge ($x=0$) of a spinful superconductor with (a) pure $p$-wave pairing and (b) $p$-wave pairing that dominates in the bulk but competes with a subleading $s$-wave instability at the boundary. The wave functions of the MBS of the spinless Kitaev chain with homogeneous $p$-wave order parameter and spatially varying $p$-wave order parameter are shown and compared to $\text{const.}\times e^{-x/\xi_t}$ in (c) and (d), respectively.}
\label{fig:MBSComp}
\end{figure}

\subsection{Consequences for MBSs}
\label{ImpactOnMBSsWF}
For the simple band structure assumed, with only two Fermi points, see \equref{Hopping}, the system is a topological class-DIII superconductor for triplet pairing and in a topologically trivial phase for singlet superconductivity\cite{FuBergProposal,Sato,QiInvar}. In the case without competing singlet pairing ($V_s = 0$), we thus obtain (Kramers pairs of) MBSs localized at the edges of the system, see \figref{fig:MBSComp}(a).

Despite being irrelevant for the topological properties of the bulk phase, the presence of a competing singlet instability has a prominent influence on the location of the MBS wave function $\phi_{\text{maj}}$.
For competing order parameters, $V^t > V^s \neq 0$, its position is shifted from the boundary towards the bulk on a scale of the coherence length $\xi_t = 1/\Delta^t$ as can be seen in \figref{fig:MBSComp}(b). This shift can be explained by the local structure of the multi-component order parameters (see \figref{fig:CompetingOP}): Due to the emergence of an additional, topologically trivial, region of predominantly singlet pairing close to the boundary, the transition point from the topologically non-trivial to the topologically trivial region now constitutes the new boundary at which the MBSs arise. Thus, the wave function is shifted into the bulk on a scale set by the two respective coherence lengths, $\xi_s$ and $\xi_t$.

We note that the phase competition between singlet (trivial) and triplet (topological) superconductivity and the associated shift of the MBSs cannot occur in spinless models, such as the Kitaev chain, as the lack of spin degrees of freedom only allows for $p$-wave pairing. The sole modification as compared to just taking spatially constant $p$-wave pairing is a slight deformation of the MBS wave function as can be seen in \figref{fig:MBSComp}(c) and (d):
For constant $p$-wave pairing, \figref{fig:MBSComp}(c), the wave function of a MBS decays exponentially from the boundary towards the bulk. The local suppression of the $p$-wave order parameter in the self-consistent treatement of a spinless system with $p$-wave pairing interaction, \figref{fig:MBSComp}(d), modifies the exponential functional form of the wave function close to the boundary. The maximum, however, is still at the boundary.
A similar effect can be seen in the spatial form of the MBS wave function in the spinful case, \figref{fig:MBSComp}(a), without competing $s$-wave pairing.

\subsection{Behavior at impurities}\label{BehaviorAroundImpur}
In addition to the suppression of triplet pairing and the emergence of a local singlet phase in a spinful superconductor, we have also identified significant Friedel oscillations in the singlet channel in the vicinity of the edge (see \figref{fig:CompetingOP}), which are absent in the triplet component. As one might expect, very similar behavior can be found in the vicinity of impurities as we will discuss next.

To analyze impurities, we impose PBC on our system and separately consider $s$-wave and $p$-wave pairing in presence of either a single non-magnetic ($\hat{U}\propto \hat{\sigma}^0$) and magnetic ($\hat{U}\propto \hat{\sigma}^z$) defect. Note that Friedel oscillations in the $s$-wave singlet order parameter in the presence of a single non-magnetic defect have already been studied by \citet{WalkerOsc}. We will extend their analysis by including $p$-wave triplet pairing and by studying magnetic impurities.

\begin{figure}[tb]
\includegraphics[scale=0.15]{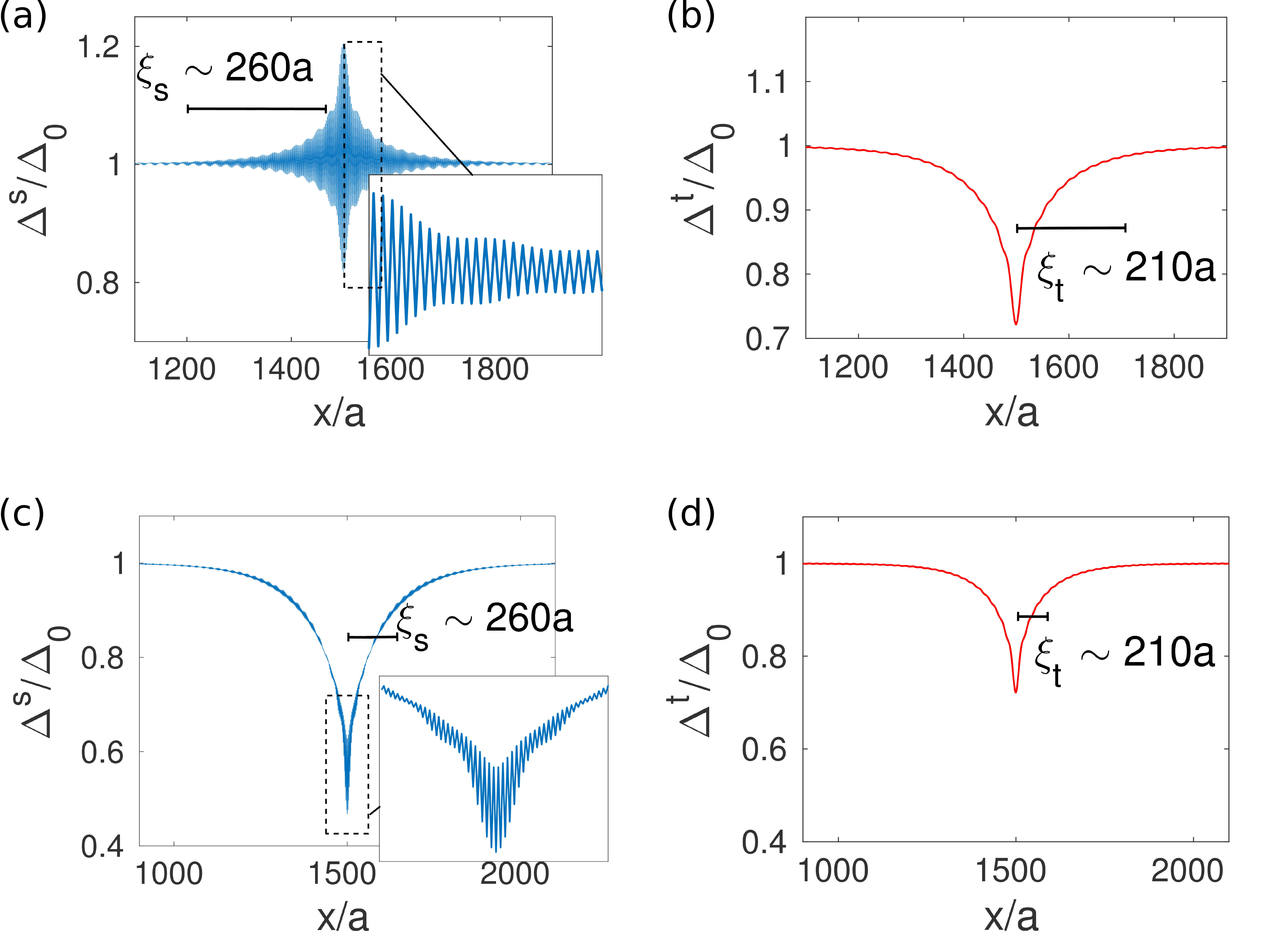}
\caption{Spatial structure of the (a) pure $s$-wave singlet ($V^s = 0.36$) and (b) pure $p$-wave triplet ($V^t=0.38$) order parameter normalized to the respective bulk value $\Delta_0^{\eta}$, in the region close to a non-magnetic impurity at site $i_0=1500a$ for a system with $L=3000$ sites, chemical potential $\mu=0$, coherence length $\xi_{\eta} = 1/\Delta_0$, impurity strength $u_0=0.5t$, and periodic boundary conditions. Part (c) and (d) show the analogous plots for a magnetic impurity.}
\label{fig:1DIMP}
\end{figure}

For a non-magnetic impurity the local structure of the $s$-wave order parameter (see \figref{fig:1DIMP}(a)) exhibits pronounced order-parameter-enhancing Friedel oscillations close to the impurity site, similar to the oscillations present close to a boundary, as considered above. The oscillations arise in the vicinity of the defect and decay towards the bulk value on the scale of the coherence length $\xi_s$, while the oscillation period is determined by $k_F$.

As in the case near the sample edge, the $p$-wave order parameter does not possess pronounced Friedel oscillations near defects, but exhibits a suppression of the superconducting gap close to the defect site (see \figref{fig:1DIMP}(b)). The order parameter recovers to its bulk value over the coherence length $\xi_t$. 

Very similar behavior is found for $s$-wave pairing in the presence of a magnetic impurity.
As can be seen in \figref{fig:1DIMP}(c), the overall behavior is governed by a dominant suppression close to the impurity site, recovering the bulk value over the coherence length $\xi_s$. Within the region of suppression, $\Delta^s$ exhibits minor oscillatory features, which are qualitatively distinct from the Friedel oscillations for the non-magnetic defect; the oscillations in the vicinity of a magnetic defect do not lead to a local enhancement of the $s$-wave gap as compared to its bulk value. 

From the local structure of the $p$-wave order parameter in \figref{fig:1DIMP}(d) it is evident that no oscillations are present. Similar to the non-magnetic defect a magnetic defect leads to a suppression of the gap in the vicinity of the impurity, which recovers on the scale of $\xi_t$. Comparing the magnitude of the suppression for a magnetic defect for the two respective pairing symmetries we find a stronger suppression of the $s$-wave gap, in accordance with the greater vulnerability of the singlet $s$-wave superconductor to magnetic impurities.

\section{Analytical understanding of Friedel oscillations}
\label{sec:origin}
In this section, we will derive an analytical expression for the impact of an isolated impurity on the superconducting order parameter in leading order in the impurity strength. From the structure and symmetry properties of this expression, the numerical observation of Friedel oscillations in the spatial structure of the singlet $s$-wave order parameter and the lack thereof in the $p$-wave triplet component are readily seen.

\subsection{Perturbative expression for spatial variations of order parameter}
We start by transforming the Hamiltonian (\ref{Hamiltonian}) with a single impurity localized at the origin, $U_i^{\sigma,\sigma'} = \delta_{i,0}U_{\sigma,\sigma'}$ in \equref{ImpurityHam}, to momentum space,
\begin{align}\begin{split}
\hat{\mathcal{H}}_{\text{0}} &= \sum_{\vec{k},\sigma} \epsilon_{\vec{k}} c_{\vec{k}\sigma}^{\dagger}  c_{\vec{k}\sigma}^{\pdagger}, \\
\hat{\mathcal{H}}_{\text{imp}} &= \sum_{\vec{k},\vec{k}'}\sum_{\sigma,\sigma'} c_{\vec{k}\sigma}^{\dagger} U_{\sigma\sigma'} c_{\vec{k}'\sigma'}^{\pdagger},\\
\hat{\mathcal{H}}_{\text{pairing}}&=V \sum_{\vec{k},\vec{k}',\vec{q}}\sum_{\sigma,\sigma',\tilde{\sigma},\tilde{\sigma}'} c_{\vec{k}+\frac{\vec{q}}{2}\sigma}^{\dagger} \chi_{\sigma\sigma'}^{\dagger}(\vec{k})c_{-\vec{k}+\frac{\vec{q}}{2}\sigma'}^{\dagger}\\ &\qquad\quad \times c_{-\vec{k}'+\frac{\vec{q}}{2}\tilde{\sigma}}\chi_{\tilde{\sigma}\tilde{\sigma}'}(\vec{k}')c_{\vec{k}'+\frac{\vec{q}}{2}\tilde{\sigma}'}
\end{split}\end{align}
with $c_{\vec{k}\sigma}$ ($c_{\vec{k}\sigma}^{\dagger}$) denoting the fermionic annihilation (creation) operators in momentum space and single-particle dispersion $\epsilon_{\vec{k}}$.  The pairing symmetries for singlet and triplet superconductivity are encoded in the momentum- and spin-dependent matrix $\chi(\vec{k})$, with relative and center-of-mass momenta denoted by $\vec{k}$ and $\vec{q}$, respectively.

A mean-field decomposition in the Cooper-channel yields
\begin{align}\label{eq:Hmfor}\begin{split}
&\hat{\mathcal{H}}_{\text{MF}} = \sum_{\vec{k},\vec{k}'}\sum_{\sigma,\sigma'} c_{\vec{k}\sigma}^{\dagger} h_{\vec{k}\sigma,\vec{k}'\sigma'} c_{\vec{k}'\sigma'}^{\pdagger} \\ &+  \sum_{\vec{k},\vec{q}}\left( \Delta(\vec{q}) c_{\vec{k}+\frac{\vec{q}}{2}\sigma}^{\dagger} \chi_{\sigma\sigma'}^{\dagger} (\vec{k})c^{\dagger}_{-\vec{k}+\frac{\vec{q}}{2}\sigma'} + \text{H.c.}\right), \\
\\
\\
\end{split}\end{align}
where we have defined $h_{\vec{k}\sigma,\vec{k}'\sigma'}=\delta_{\vec{k},\vec{k}'}\delta_{\sigma,\sigma'}\epsilon_{\vec{k}}   +  U_{\sigma\sigma'}$ and the gap function
\begin{equation}\label{eq:selfconk}
\Delta(\vec{q}) = V\sum_{\vec{k}} \chi_{\sigma\sigma'}(\vec{k}) \langle c_{-\vec{k}+\frac{\vec{q}}{2}\sigma}\, c_{\vec{k}+\frac{\vec{q}}{2}\sigma'}\rangle \text{.}
\end{equation}
Introducing the spinor $\hat{\Psi}_{\vec{k}\sigma} = \left( c_{\vec{k}\sigma},c_{-\vec{k}\sigma}^{\dagger}\right)^{T}$, we rewrite the Hamiltonian in \equref{eq:Hmfor} in the standard BdG form:

\begin{widetext}
\begin{equation}
\hat{\mathcal{H}}_{\text{MF}} = \sum_{\vec{k},\vec{k}'} \hat{\Psi}_{\vec{k}\sigma}^{\dagger} \hat{\mathcal{H}}_{\vec{k}\sigma,\vec{k}'\sigma'}^{\text{BdG}} \hat{\Psi}_{\vec{k}'\sigma'}, \quad \hat{\mathcal{H}}_{\vec{k}\sigma,\vec{k}'\sigma'}^{\text{BdG}}= \begin{pmatrix}
h_{\vec{k}\sigma,\vec{k}'\sigma'} & \Delta(\vec{k}-\vec{k}')\chi_{\sigma\sigma'}^{\dagger}(\frac{\vec{k}+\vec{k}'}{2})\\
\Delta^*(\vec{k}'-\vec{k})\chi_{\sigma\sigma'}(\frac{\vec{k}+\vec{k}'}{2}) & -h_{-\vec{k}'\sigma',-\vec{k}\sigma}
\end{pmatrix}\text{.}
\end{equation}

Denoting the Nambu Green's function by $\mathcal{G}_{\sigma\sigma'}(\omega_n;\vec{k},\vec{k}') = \left(i\omega_n-\hat{\mathcal{H}}^{\text{BdG}}\right)_{\vec{k}\sigma,\vec{k}'\sigma'}^{-1}$ with $\omega_n$ representing fermionic Matsubara frequencies,  \equref{eq:selfconk} is rewritten as
\begin{equation}\label{eq:sc1}
\Delta(\vec{q}) = VT\sum_{\omega_n} \sum_{\vec{k}} \chi_{\sigma\sigma'}(\vec{k})\left( \mathcal{G}_{\sigma\sigma'}(\omega_n;\vec{k}-\vec{q}/2,\vec{k}+\vec{q}/2)\right)_{1,2}.
\end{equation}
Assuming the impurity to be weak,  we only keep terms linear in $\hat{U}$ and neglect the spatial variations\cite{WalkerOsc} of the gap on the right-hand side of \equref{eq:sc1}, $\Delta(\vec{q}) \rightarrow \Delta_0$ with  $\Delta_0$ denoting the solution of the homogeneous gap equation in the absence of the impurity. Furthermore, we linearize \equref{eq:sc1} in the superconducting order parameter which is valid in the vicinity of $T_c$. 
With these approximations, \equref{eq:sc1} can be written as 
\begin{equation}\label{eq:Fq}
F(\vec{q}) = V\Delta_0 T\sum_{\omega_n}\sum_{\vec{k}} \frac{\epsilon_{\vec{k}+\frac{\vec{q}}{2}}}{(\omega_{n}^{2}+\epsilon_{\vec{k}+\frac{\vec{q}}{2}}^{2})(\omega_{n}^{2}+\epsilon_{\vec{k}-\frac{\vec{q}}{2}}^{2})} \operatorname{tr}\left[\chi(\vec{k})\,\hat{U}\,\chi^{*}(\vec{k}-\vec{q}/2)\right],
\end{equation}
\end{widetext}
with $F(\vec{q}) \equiv \Delta (\vec{q}) - \delta_{\vec{q},0}\Delta_{0}$ describing the impurity-induced alteration of the superconducting gap and $\text{tr}$ denoting the trace in spin-space.
As we will see below, the crucial qualitative differences in the behavior of the superconducting phases in the presence isolated impurities are captured by \equref{eq:Fq}. We thus conclude that the Born approximation applied above suffices to describe the relevant physics.

\subsection{Non-magnetic impurities in 1D}
First, we study the general expression in \equref{eq:Fq} for a non-magnetic impurity of the form $\hat{U}=u_0\hat{\sigma}^0$ and consider the cases of pure $s$-wave and pure $p$-wave pairing separately.

\begin{figure}[t]
\centering \includegraphics[width =\linewidth]{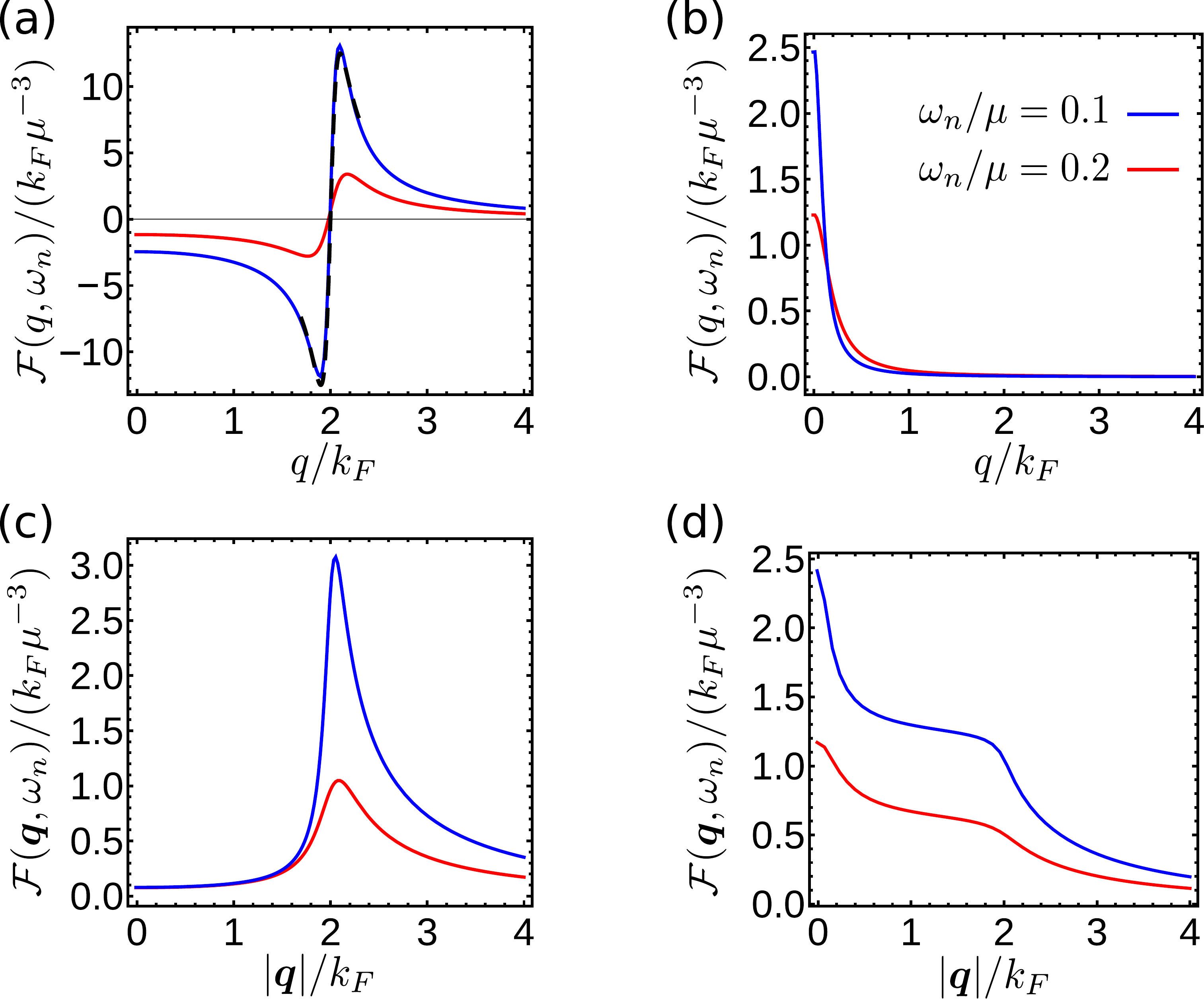}
\caption{Momentum dependence of $\mathcal{F}$ at two different frequencies, $\omega_n/\mu=0.1$ (blue solid line) and $\omega_n/\mu=0.2$ (red solid line), for the four different cases of $s$-wave in 1D (a),  $p$-wave in 1D (b), $s$-wave in 2D (c), and $p$-wave in 2D (d). The dashed black line in (a) refers to the approximate expression in \equref{eq:Fs1d}.}
\label{AnalyticCurves}
\end{figure}

We take the singlet pairing interaction to be \mbox{$\chi_{s}(\vec{k}) = i\hat{\sigma}^{y}$}, i.e.~momentum independent. With these approximations, the expression in \equref{eq:Fq} reduces to
\begin{equation}
 F(q) = -V\Delta_0 u_0 L^d T  \sum_{\omega_n} \mathcal{F}(q,\omega_n), \label{FcalIntro}
 \end{equation}
with system length $L$, $d=1$, and
\begin{equation}
 \mathcal{F}(q,\omega_n) = 2\int\limits_{-\infty}^{\infty} \frac{dk}{2\pi} \frac{\epsilon_{k+\frac{q}{2}}}{(\omega_{n}^{2}+\epsilon_{k+\frac{q}{2}}^2)(\omega_{n}^{2}+\epsilon_{k-\frac{q}{2}}^2)}. \label{1DNonmagswave}
\end{equation}
The graph in \figref{AnalyticCurves}(a) shows plots of $\mathcal{F}$ for different $\omega_n$, taking $\epsilon_{k}=\frac{k}{2m}-\mu$ for concreteness. Due to the symmetry of $\mathcal{F}$, i.e. $\mathcal{F}(q,\omega_n) = \mathcal{F}(-q,\omega_n)$, only the region $q > 0$ is shown. A distinct peak around $q=2k_F$ is visible in the graph, which becomes sharper as $\omega_n$ decreases. We expand \equref{1DNonmagswave} around $q=2k_F +\delta q$, with $|\delta q| \ll k_F$, and furthermore linearize $\epsilon_{k\pm q/2} \sim \pm v_F (k\pm \delta q/2)$ (expecting the main contribution to arise from the Fermi surface). This leads to
\begin{align}\label{eq:Fs1d}
\mathcal{F}(q,\omega_n) &\sim \int\limits_{-\infty}^{\infty} \frac{dk}{2\pi} \frac{v_F \, \delta q}{(\omega_n^2 + v_F^2\, (k+\frac{\delta q}{2})^2)(\omega_n^2 + v_F^2 \, (k-\frac{\delta q}{2})^2)}\nonumber \\&= \frac{1}{|\omega_n|}\frac{\delta q}{(2\omega_n)^2 + (v_F \, \delta q)^2}
\end{align}
and the pole-structure of $\mathcal{F}(q,\omega_n)$ at $q=\pm 2k_F \pm i2\omega_n/v_F$ reproduces the oscillation period equaling $2k_F$ in the numerical results in \secref{1DResults}.

For the case of $p$-wave pairing, the interaction is chosen to be $\chi_{t}(k) = i\hat{\sigma}^{y}\hat{\sigma}^{y} k/k_F = i\hat{\sigma}^{0} k/k_F$. Substituting this into \equref{eq:Fq} yields
\begin{align}
F(q) &= V \Delta_0 u_0 L^d T \sum_{\omega_n} \mathcal{F}(q,\omega_n), \quad \text{with} \label{eq:F1}\\ \mathcal{F}(q,\omega_n) &= 2\int\limits_{-\infty}^{\infty} \frac{dk}{2\pi} \frac{\epsilon_{k+\frac{q}{2}} k(k-\frac{q}{2})/k_F^2 }{(\omega_{n}^{2}+\epsilon_{k+\frac{q}{2}}^2)(\omega_{n}^{2}+\epsilon_{k-\frac{q}{2}}^2)} \label{eq:F2}.
\end{align}
In contrast to $s$-wave pairing, $\mathcal{F}(q,\omega_n)$ does not exhibit a peak around $q=2k_F$, as can be seen in \figref{AnalyticCurves}(b). No oscillations are, thus, expected in accordance with our numerical results for the local structure of the $p$-wave superconducting order parameter. 
This can be qualitatively understood as follows. As in the case of $s$-wave pairing, for $q\simeq 2k_F$, the weight of the poles of the denominator of the integrand in \equref{eq:F2} is greatest for $k\simeq 0$. 
However, the $p$-wave nature of the pairing interaction, represented by the factor $k$ in the numerator, leads to a cancellation by a simultaneously vanishing numerator. This can be more explicitly seen by performing an approximation analogous to the $s$-wave case. Linearising the dispersion relation, i.e.~$\epsilon_{k\pm q/2}=\pm v_F (k\pm \delta q/2)$, and expanding about $q=2k_F + \delta q$, the expression given in \equref{eq:F2} becomes independent of $\delta q$, and, hence, does not exhibit the pole structure as in \equref{eq:Fs1d}. This explains why no Friedel-like oscillations are found for $p$-wave pairing.

\subsection{Non-magnetic impurities in 2D}
So far, we have been focusing on the 1D case. As the analytical result in \equref{eq:Fq} can be readily applied to higher dimensions as well, let us next discuss its predictions for Friedel-like oscillations around defects in two spatial dimensions. In \secref{sec:roledim} below, the numerical solutions of the 2D inhomogeneous gap equations in the presence of impurities will be presented.

For a 2D system with $s$-wave pairing interaction (choosing $\chi_{s}(\vec{k}) = i\hat{\sigma}^{y}$ as before), we obtain
\begin{equation}\label{eq:Fs2}
\mathcal{F}(\vec{q},\omega_n) = 2\int \frac{d^2 \vec{k}}{(2\pi)^2} \frac{\epsilon_{\vec{k}+\frac{\vec{q}}{2}} }{(\omega_n^2 + \epsilon_{\vec{k}+\frac{\vec{q}}{2}}^2)(\omega_n^2 + \epsilon_{\vec{k}-\frac{\vec{q}}{2}}^2)} 
\end{equation}
in \equref{FcalIntro} with $d=2$.
Due to the rotational symmetry of the dispersion $\epsilon_{\vec{k}}=\frac{\vec{k}^2}{2m}-\mu$, the expression only depends on $|\vec{q}|$, $\mathcal{F}(\vec{q},\omega_n) = \mathcal{F}(|\vec{q}|\vec{e}_{x},\omega_n)$. Based on the presence of a peak around $|\vec{q}|\simeq 2k_F$, see \figref{AnalyticCurves}(c), we expect Friedel-like oscillations in the $s$-wave order parameter, similar to the 1D case.

For a general triplet vector, $\chi_{t}(\vec{k})= \vec{d}(\vec{k}) \cdot \vec{\sigma} i\hat{\sigma}^{y}$, we find
\begin{equation}
\mathcal{F} = 2 \int \frac{d^2 \vec{k}}{(2\pi)^2}\,\,  \frac{\epsilon_{\vec{k}+\frac{\vec{q}}{2}} \, \vec{d}(\vec{k}) \cdot \vec{d}^*(\vec{k}-\frac{\vec{q}}{2}) }{(\omega_n^2 + \epsilon_{\vec{k}+\frac{\vec{q}}{2}}^2)(\omega_n^2 + \epsilon_{\vec{k}-\frac{\vec{q}}{2}}^2)} 
\end{equation}
in \equref{eq:F1} with $d=2$.
For concreteness, let us further assume a $d$-vector of the form $\vec{d}(\vec{k}) = (k_y,-k_x,0)^T/k_F$, leading to
\begin{equation}\label{eq:Fp2}
\mathcal{F} = 2\int \frac{d^2 \vec{k}}{(2\pi)^2} \,\,  \frac{\epsilon_{\vec{k}+\frac{\vec{q}}{2}} \,\, \vec{k}(\vec{k}-\frac{\vec{q}}{2})/k_F^2 }{(\omega_n^2 + \epsilon_{\vec{k}+\frac{\vec{q}}{2}}^2)(\omega_n^2 + \epsilon_{\vec{k}-\frac{\vec{q}}{2}}^2)} \text{.}
\end{equation}
Again, $\mathcal{F}(\vec{q},\omega_n)$ is rotationally symmetric, such that $\mathcal{F}(\vec{q},\omega_n) = \mathcal{F}(|\vec{q}|\vec{e}_{x},\omega_n)$. The numerical evaluation of \equref{eq:Fp2}, as shown in \figref{AnalyticCurves}(d), does not exhibit a pronounced peak around $|\vec{q}|\simeq 2k_F$, but does show an enhancement of $\mathcal{F}$ for $|\vec{q}| < 2k_F$. This is due to the continuum of $\vec{q}$-vectors connecting parts of the Fermi surface. 

Consequently, we expect the difference in the behavior of $s$- and $p$-wave order parameters around non-magnetic impurities to be smaller in higher dimensions than in 1D; in accordance with our numerical results presented in \secref{sec:roledim} below, Friedel-like oscillations are expected in both $s$- and $p$-wave components in higher dimensions, whereas, in 1D, only the $s$-wave channel shows oscillatory behavior.

\subsection{Magnetic impurities}\label{Magnetic1D2DAnal}
Let us next analyze the predictions of our analytical expression in \equref{eq:Fq} for the case of a magnetic impurity, i.e., for $\hat{U} = \vec{S}\cdot \vec{\sigma}$ with some real vector $\vec{S}$.

Inserting a general singlet-pairing function, $\chi(\vec{k})=i\hat{\sigma}^{y} f(\vec{k})$ with $f(\vec{k}) = f(-\vec{k})$, the spin-trace in \equref{eq:Fq} vanishes, leading to $F=0$. Consequently, there is no impact on the local superconducting order parameter within the leading-order perturbative treatment presented in this section. The suppression of the superconducting order parameter, including the minor oscillatory features, present in the local structure of the singlet $s$-wave (see \figref{fig:1DIMP}(c)) are not captured by the analytical calculation presented here as a result of the approximations made to derive the expression in \equref{eq:Fq}. 

Taking a general triplet term, $\chi(\vec{k}) = (\vec{d}(\vec{k})\cdot \vec{\sigma}) i\hat{\sigma}^{y}$ with triplet vector satisfying $\vec{d}(\vec{k}) = -\vec{d}(-\vec{k})$, \equref{eq:Fq} assumes the form
\begin{align}\begin{split}
F(\vec{q}) &= 2i V\Delta_0 T \\ &\times \sum_{\omega_n}\sum_{\vec{k}} \frac{\vec{S}\cdot \left(\vec{d}(\vec{k}) \times \vec{d}^*(\vec{k}-\frac{\vec{q}}{2})\right) \epsilon_{\vec{k}+\frac{\vec{q}}{2}}}{(\omega_{n}^{2}+\epsilon_{\vec{k}+\frac{\vec{q}}{2}}^{2})(\omega_{n}^{2}+\epsilon_{\vec{k}-\frac{\vec{q}}{2}}^{2})}. \label{TripletExprFq}
\end{split}\end{align}
Interestingly, the resulting leading order correction is purely imaginary for real-valued triplet vectors. This means that the phase of the order parameter changes in the vicinity of the impurity and, thus, induces a time-reversal-symmetry-breaking component in the superconducting order parameter. In the 1D case we have studied numerically in \secref{BehaviorAroundImpur}, we did not obtain this phase as the triplet vector $\vec{d}(\vec{k})$ was colinear leading to $\vec{d}(\vec{k}) \times \vec{d}^*(\vec{k}-\frac{\vec{q}}{2})=0$ in \equref{TripletExprFq}. When investigating the 2D case numerically in the next section, we will indeed recover the emergence of a local imaginary part.

We finally note that there are crucial differences to the local phase shift by $\pi$ in $s$-wave superconductors in the vicinity of magnetic impurities studied, e.g., in \refscite{PhysRevB.95.104521,PhysRevB.92.064503,PhysRevB.93.165418}. First, the local phase shift in the $p$-wave superconductor is continuous and incommensurate with $\pi$, thus breaking time-reversal-symmetry in the superconducting order parameter itself. Secondly, it occurs already at linear order in the impurity strength and as such does not require the impurity potential to exceed a certain strength.

\section{Numerical results for higher dimensions}
\label{sec:roledim}
In addition to the numerical treatment of 1D systems and the analytic approach in both dimensions, we follow up with a numerical examination of 2D square lattices with $L\times L$ sites and the influence of inversion symmetry breaking at impurities and boundaries on the different superconducting order parameters. 

\begin{figure}[tb]
\includegraphics[width=\linewidth]{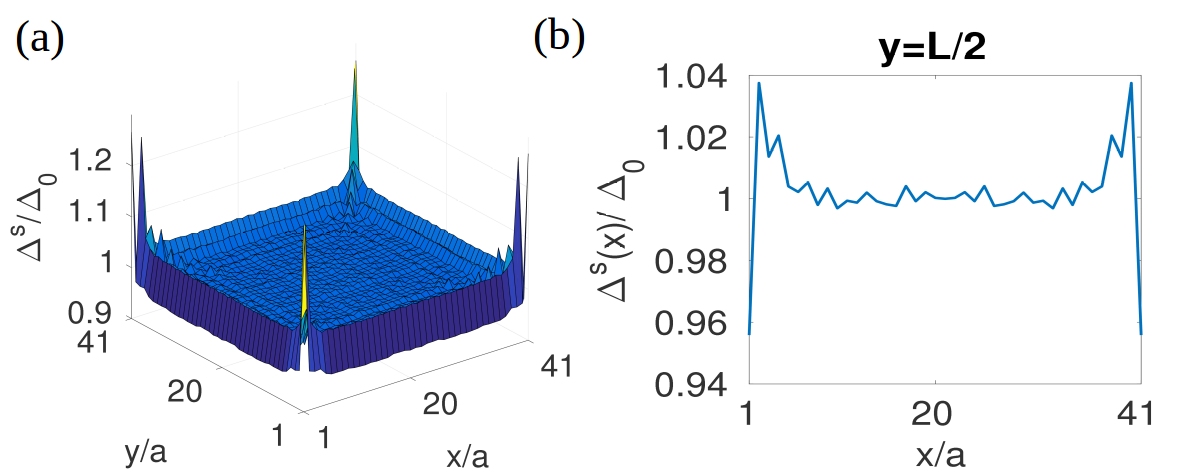}
\caption{(a) Surface plot of $\Delta^s$ for a system of $41\times 41$ sites with OBC, coupling constant $V^s=0.5$, energy cut-off $\epsilon_c=0.7t$  and chemical potential $\mu=0t$. (b) $\Delta^s$ at $y=L/2$ along $x$.}
\label{fig:2DsOBC}
\end{figure}

The result of the self-consistent calculation for a 2D square lattice with OBC and pure $s$-wave pairing is depicted in \figref{fig:2DsOBC}.
Similar to the one dimensional case, the order parameter exhibits oscillations along with an enhancement in the vicinity of the edges and at the corners, though the effect is weaker than in 1D.
\begin{figure}[tb]
\includegraphics[width=\linewidth]{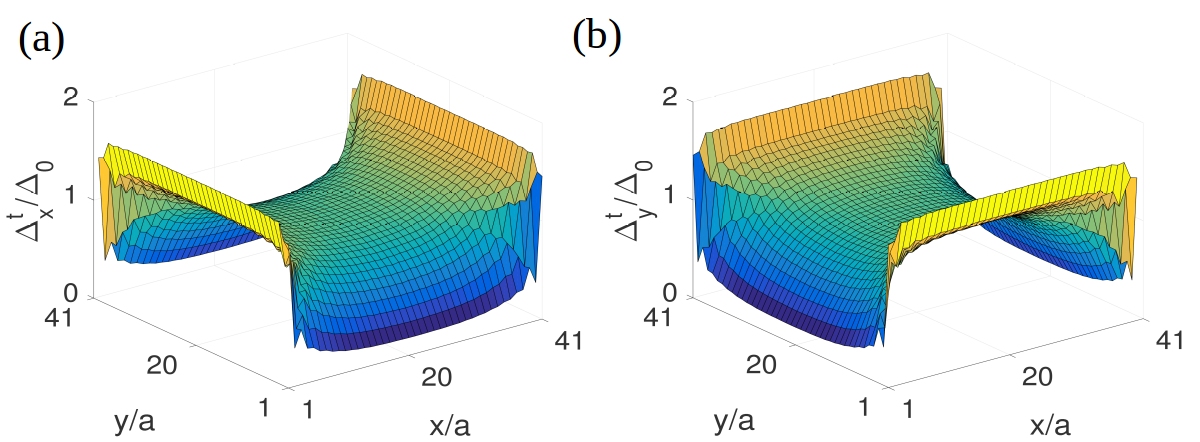}
\caption{(a) $\Delta^t_x$ and (b) $\Delta^t_y$ for a system of $41\times 41$ sites with OBC in $x$ and $y$, coupling constant $V^t=1.2$, energy cut-off $\epsilon_c=1.1t$ and chemical potential $\mu=0$.}
\label{fig:2DpOBC}
\end{figure}

We have also studied the behavior of 2D $p$-wave pairing in the vicinity of edges taking a bulk triplet vector of the form $\vec{d} = (\sin(k_y),-\sin(k_x),0)^T$. In contrast to the $s$-wave case, the two components $\Delta^t_x$ and $\Delta^t_y$ of the triplet order parameter transforming as $k_y$ and $k_x$, respectively, do not exhibit a spatial oscillation, but a smooth decrease towards one boundary and a smooth increase towards the other, see \figref{fig:2DpOBC}.

\begin{figure}[tb]
\includegraphics[width=\linewidth]{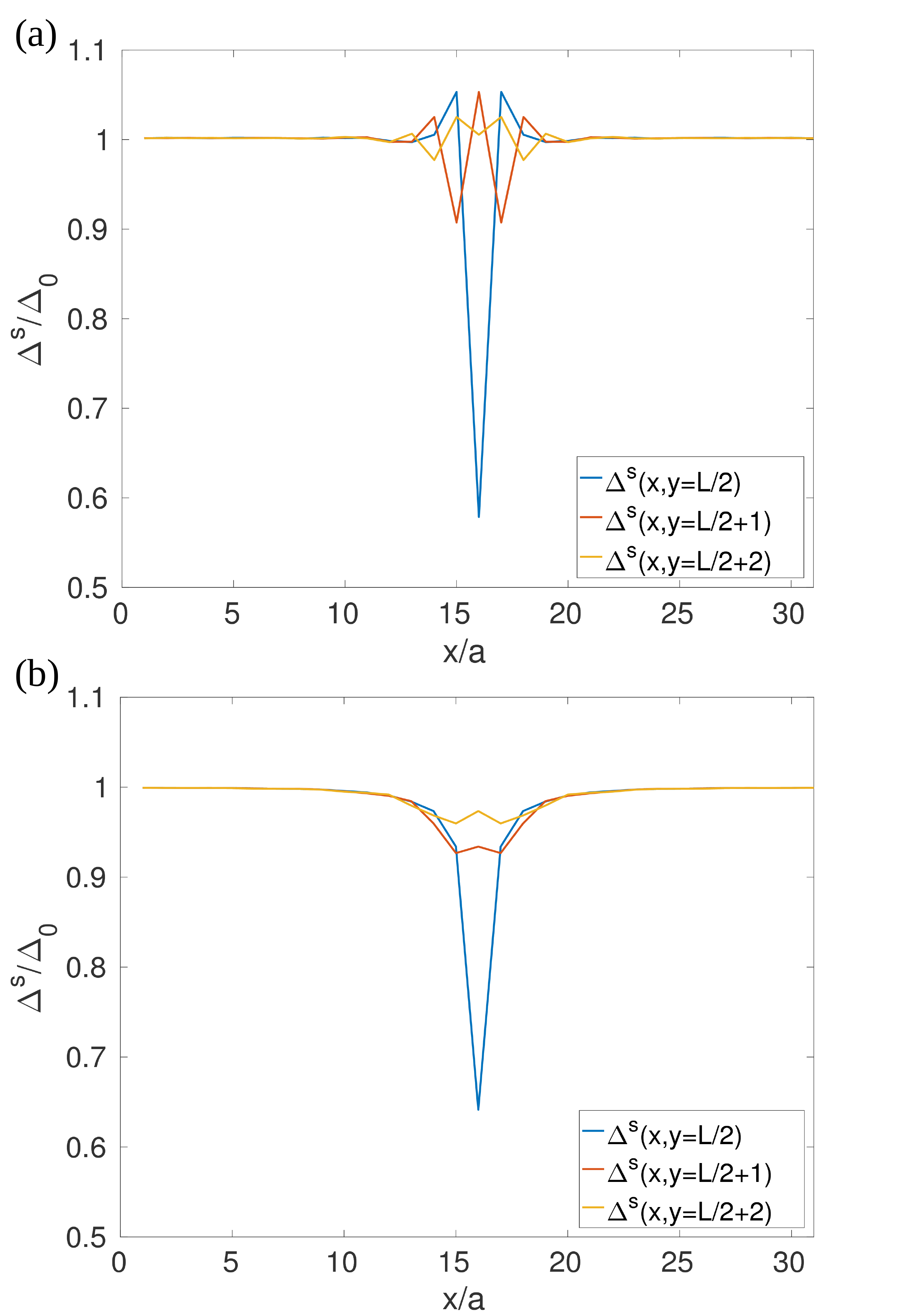}
\caption{Cross sections of $\Delta^s$ for a system of $31\times 31$ sites and PBC, coupling constant $V^s=0.5$, energy cut-off $\epsilon_c=0.7t$ and chemical potential $\mu=-0.3t$ with (a) non-magnetic and (b) magnetic impurities  of strength $u_0=0.25t$ located at $(L/2,L/2)$.}
\label{fig:2DsIMP}
\end{figure}

The influence of non-magnetic and magnetic defects on the spatial structure of the 2D $s$-wave order parameter closely resembles that of the 1D case (see \figref{fig:2DsIMP}). A non-magnetic defect leads to enhanced oscillations close to the defect site, while a magnetic inhomogeneity leads to a dominant suppression. This behavior is in good accordance with our analytic treatment in \secref{sec:origin}.

\begin{figure}[t]
\includegraphics[width=\linewidth]{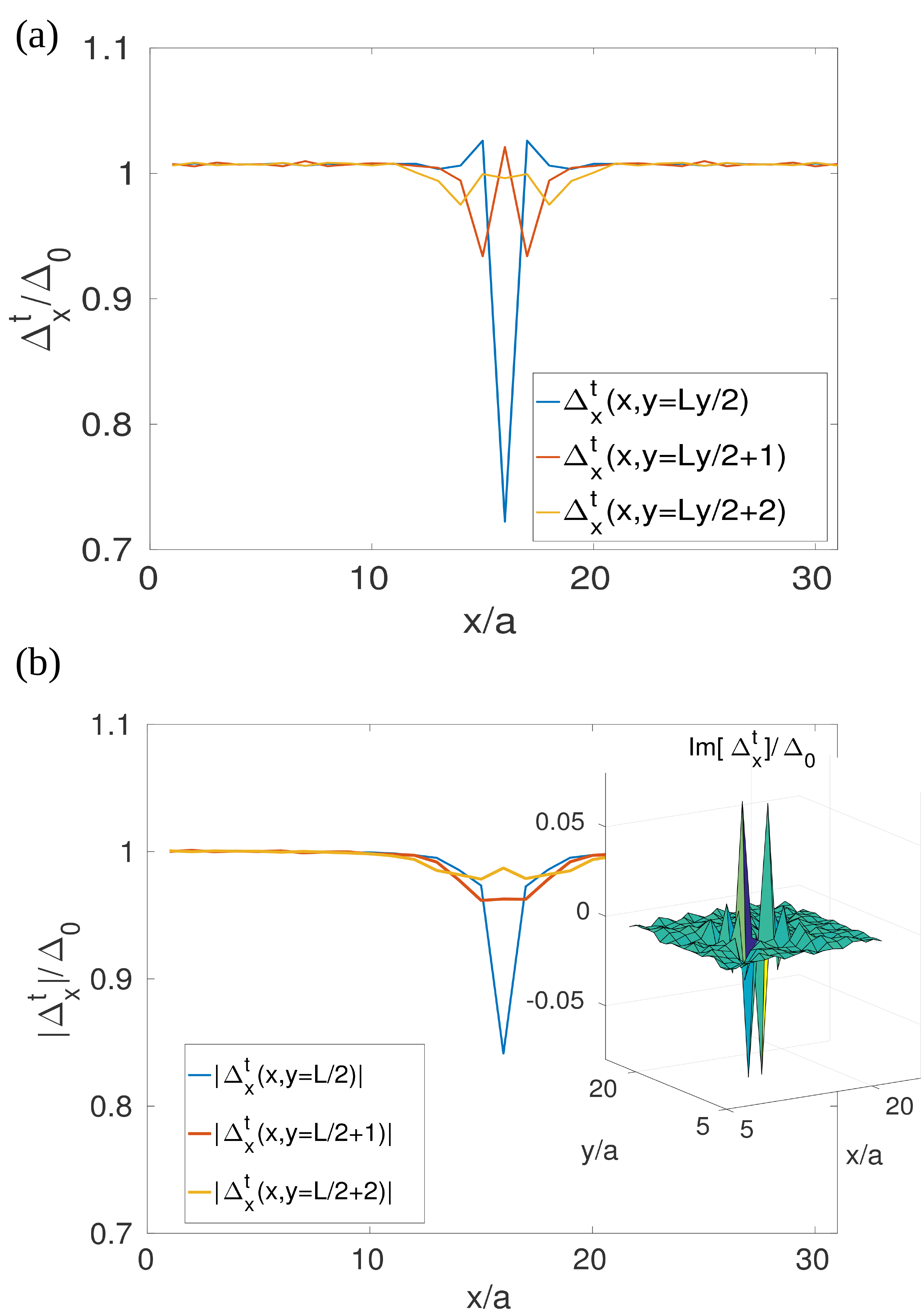}
\caption{Same as \figref{fig:2DsIMP}, for triplet pairing using $V^t=0.7$ and energy cut-off $\epsilon_c=1.1t$. In the case of the magnetic impurity in part (b), the order parameter becomes complex in the vicinity of the defect as anticipated by our analytical analysis in \secref{Magnetic1D2DAnal}. For this reason, we show the absolute value $|\Delta_x^t|$ and $\Im[\Delta_x^t]$ (inset) in part (b).}
\label{fig:DpIMPcut}
\end{figure}

A comparison of these results with the effect of non-magnetic and magnetic defects on the $p$-wave order parameter only yields minor differences (see \figref{fig:DpIMPcut}). While a magnetic defect leads to a suppression of the order parameter and a local phase, in agreement with our results in \secref{sec:origin}, a non-magnetic defect leads to an oscillatory enhancement of the two order parameter components close to the impurity. This limits the discernibility of $s$-wave and $p$-wave by defect-induced spatial variations. The only sharp qualitative difference in the behavior of $s$- and $p$-wave pairing in the vicinity of defects is that only the latter exhibits a local change of phase in the case of a magnetic impurity.

\section{Detection by JSTM}
\label{sec:observe}
As a promising experimental tool to detect the spatial textures of competing superconducting order parameters and of MBS wave functions, we discuss JSTM\cite{MartinProp,FirstExperimentJSTM,FabricationOfTips,Hamidian:2016aa,Yazdani,SuperconductingSTM} and compute the expected signatures in the Jospheson current. We extend previous theoretical work (see, in particular, \refcite{SuperconductingSTM}) to non-vanishing voltage drops across the interface and to topological superconductors with MBSs.

To this end, we assume a Hamiltonian
\begin{equation}
\hat{\mathcal{H}} = \hat{\mathcal{H}}_{sc} + \hat{\mathcal{H}}_{\text{T}} + \hat{\mathcal{H}}_{\text{tip}} \text{,}
\end{equation}
with the subsystem Hamiltonians $\hat{\mathcal{H}}_{sc}$ and $\hat{\mathcal{H}}_{\text{tip}}$ for the inhomogeneous superconductor of interest and the superconducting tip, respectively. Tunneling between the two subsystems is mediated by the tunneling Hamiltonian $\hat{\mathcal{H}}_{\text{T}}$.	The Hamiltonian of the inhomogeneous superconductor is obtained from our self-consistent calculations, while the tip Hamiltonian $\hat{\mathcal{H}}_{\text{tip}}=\hat{\mathcal{H}}_{\text{tip}}^{n} + \hat{\mathcal{H}}_{\text{tip}}^{sc}$, consisting of normal state and superconducting terms, is assumed to model an atomically sharp homogeneous superconducting tip.
The superconducting part reads as
\begin{equation}
\hat{\mathcal{H}}_{\text{tip}}^{\text{sc}}= \sum_{\sigma\sigma'} \Delta_{\text{tip}}^{\sigma\sigma'} d_{\sigma}^{\dagger}d_{\sigma'}^{\dagger} + \text{H.c.}\text{,}
\end{equation}
with operators $d_{\sigma}^{(\dagger)}$ that annihilate (create) an electron of spin $\sigma$ in the tip, and $\hat{\mathcal{H}}_{\text{tip}}^{n}$ is taken to describe a non-interacting, spin-diagonal, spectrum in the tip with density of state $\rho_0$ at the Fermi level. The Hamiltonian
\begin{equation}\label{eq:tun}
\hat{\mathcal{H}}_{\text{T}}= T_0 \sum_{\sigma} \left( c_{i_0 \sigma}^{\dagger} d^\pdagger_{\sigma} + d_{\sigma}^{\dagger}c^\pdagger_{i_0\sigma} \right) \text{,}
\end{equation}
describing tunneling (with amplitude $T_0$) between the site $i_0$ of the superconductor and the tip, is taken to preserve spin. 

Throughout this work, we take a tunneling amplitude of $T_0 = 0.01 t$ which corresponds to the weak-tunneling limit, $T_0/t\ll 1$. This allows to neglect any back action of the presence of the superconducting tip on the order parameter in the inhomogeneous superconductor and vice versa.

The Josephson current crucially depends on the difference $\Delta\phi = \phi_s - \phi_t$ of the superconducting phases of the tip ($\phi_t$) and of the inhomogeneous superconductor ($\phi_s$). These phases can be gauged away from the superconducting terms by an operator transformation $c_{i\sigma} \rightarrow c_{i\sigma}\cdot e^{i\phi_s/2}$ and \mbox{$d_{\sigma} \rightarrow d_{\sigma} \cdot e^{i\phi_t/2}$} such that they only appear via $\Delta\phi$ in the tunneling Hamiltonian 
\begin{equation}
\hat{\mathcal{H}}_{\text{T}} = T_0 \sum_{\sigma} \left( e^{i\Delta\phi/2} \, c_{i_0 \sigma}^{\dagger}d^\pdagger_{\sigma} + e^{-i\Delta\phi/2}d_{\sigma}^{\dagger}c^\pdagger_{i_0 \sigma}\right) \text{.}
\end{equation}
This gauge transformation allows us to take the order parameters $\Delta_s$ and $\Delta_t$ as real parameters below. Closely following the derivation given in \refcite{mahan} we find the Josephson tunneling current between the inhomogeneous superconductor and the superconducting tip to take the general form
\begin{widetext}\begin{subequations}\label{eq:IJ}\begin{align}\label{eq:IJ1}
I_{J}(t,V) = -2eT_0^2 I_{J,1}(V) \sin(\Delta\phi-2eVt) -2eT_0^2 I_{J,2}(V) \cos(\Delta\phi-2eVt) \text{,}
\end{align}
where $I_{J,1}(V)$ and $I_{J,2}(V)$ represent the integrals
\begin{align}
I_{J,1} &= \int \frac{d\epsilon}{\pi}\left[f(\epsilon-eV) \Re \mathcal{F}_{s,\sigma\sigma'}^{\dagger}(\epsilon+i0^+)\Im \mathcal{F}_{t,\sigma'\sigma}(\epsilon-eV+i0^+)+ f(\epsilon) \Im \mathcal{F}_{s,\sigma\sigma'}^{\dagger}(\epsilon+i0^+)\Re\mathcal{F}_{t,\sigma'\sigma}(\epsilon-eV+i0^+)\right] \label{I1Integral} \\
I_{J,2} &= \int \frac{d\epsilon}{\pi}[f(\epsilon-eV)-f(\epsilon)]\Im\mathcal{F}_{s,\sigma\sigma'}^{\dagger}(\epsilon+i0^+)\Im\mathcal{F}_{t,\sigma'\sigma}(\epsilon+i0^+-eV). \label{I2Integral}
\end{align}\end{subequations}
\end{widetext}
Here, $f(\epsilon)$ is the Fermi-distribution function, $eV$ the voltage drop across the interface, and $\mathcal{F}_s$ and $\mathcal{F}_{t}$ denote the anomalous retarded Green's functions of the inhomogeneous superconductor and the superconducting tip, respectively. Summation over spin-indices is implied. The superconducting tip is assumed to be in an isotropic $s$-wave state, such that
\begin{equation}
\mathcal{F}_{t,\sigma'\sigma}(\epsilon+i0^+) = \frac{\rho_0\pi\Delta_{\text{tip}}}{\sqrt{\Delta_{\text{tip}}^2 - (\epsilon+i0^+)^2}} (i\hat{\sigma}^y)_{\sigma'\sigma}\text{.}
\end{equation}
The sine-term in \equref{eq:IJ1} produces a dc current of electron pairs at zero voltage, while at finite voltages it leads to the oscillating currents known as Josephson effect. The cosine contribution\cite{harriscosine} in \equref{eq:IJ1} is often neglected and vanishes for $eV=0$, due to the term $[f(\epsilon-eV)-f(\epsilon)]$ in the integral in \equref{I2Integral}. We shall later see, that this contribution leads to pronounced beat signatures in the Josephson current.
\subsection{Vanishing voltage drop $eV=0$}
For a vanishing voltage drop across the interface, $eV=0$, the expression for the Josephson current in \equref{eq:IJ} reduces to 
\begin{align}\label{eq:525}
I_J&(V=0) = -2eT_0^2 \, \sin(\Delta\phi)\nonumber \\ &\times \int \frac{d\epsilon}{\pi} f(\epsilon) \Im \left[ \mathcal{F}_{s,\sigma\sigma'}^{\dagger}(\epsilon+i0^+) \mathcal{F}_{t,\sigma'\sigma}(\epsilon+i0^+)\right] \text{,}
\end{align}
as the integral associated with the cosine contribution vanishes. The critical Josephson current is given by $I_c = 2eT_0^2\int \frac{d\epsilon}{\pi} f(\epsilon) \Im \left[ \mathcal{F}_{s,\sigma\sigma'}^{\dagger}(\epsilon+i0^+) \mathcal{F}_{t,\sigma'\sigma}(\epsilon+i0^+)\right]$.
\begin{figure}[tb]
\includegraphics[width=\linewidth]{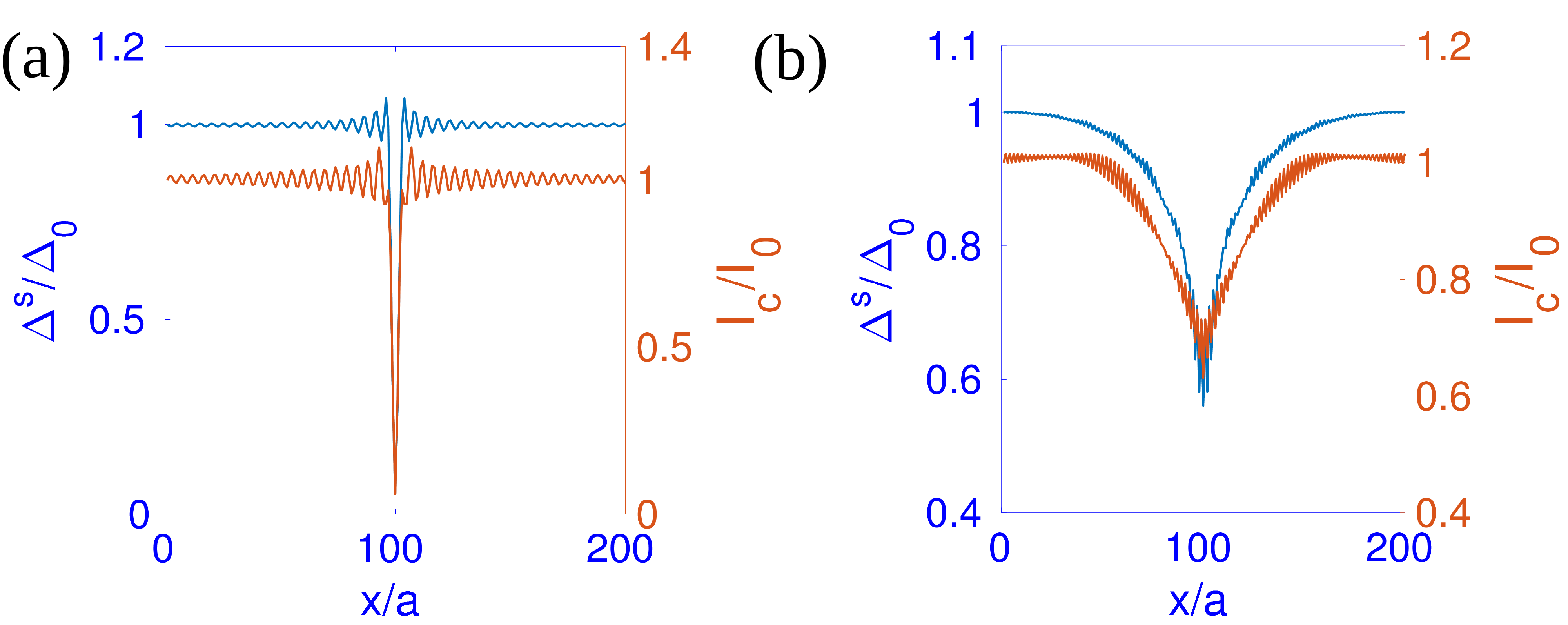}
\caption{Spatial form of the critical Josephson current $I_c$ (red), normalized to the critical current of the clean case $I_0$, and of the order parameter (blue) for (a) non-magnetic impurity at site $i_0=100a$, for $\mu_s=-t$, $V^s=0.5$, $U_{\text{imp}}=1t$, $\xi_s=14a$, $\Delta_{\text{tip}}=5\Delta_0^s$ and (b) magnetic impurity at site $i_0=100a$, for $\mu_s=0$, $V^s=0.5$, $U_{\text{imp}}=0.5t$, $\xi_s=54a$, $\Delta_{\text{tip}}=3\Delta_0^s$.}
\label{fig:IJeV0}
\end{figure}

The numerical results for the spatial variation of the critical Josephson current are presented in \figref{fig:IJeV0} for both a non-magnetic and a magnetic defect in a superconductor with $s$-wave pairing symmetry and OBC. For better optical discernibility in the graphs, the coupling constants have been chosen to lead to short coherence lengths $\xi_s$ and large oscillations periods of the Friedel oscillations (governed by $k_F$). Both the local structure of the order parameter $\Delta^s$ (blue line) and the critical Josephson current $I_c$ (red line) have been normalized to their respective bulk values, $\Delta_0$ and $I_0$, in the absence of defects. As is evident from \figref{fig:IJeV0} the Josephson current reproduces the spatial behavior of the order parameter for both magnetic and non-magnetic defects, and especially carries a signature of the Friedel oscillations for the non-magnetic defect. 

Furthermore, the minor oscillatory features present within the suppression of the order parameter close to the magnetic defect are mapped by the Josephson current. Our numerical results affirm the work presented in \refcite{SuperconductingSTM}. Next, we will extend this technique to a non-vanishing potential drop $eV\neq 0$ across the junction interface.
\subsection{Finite voltage $eV \neq 0$}
\begin{figure}[tb]
\includegraphics[width=\linewidth]{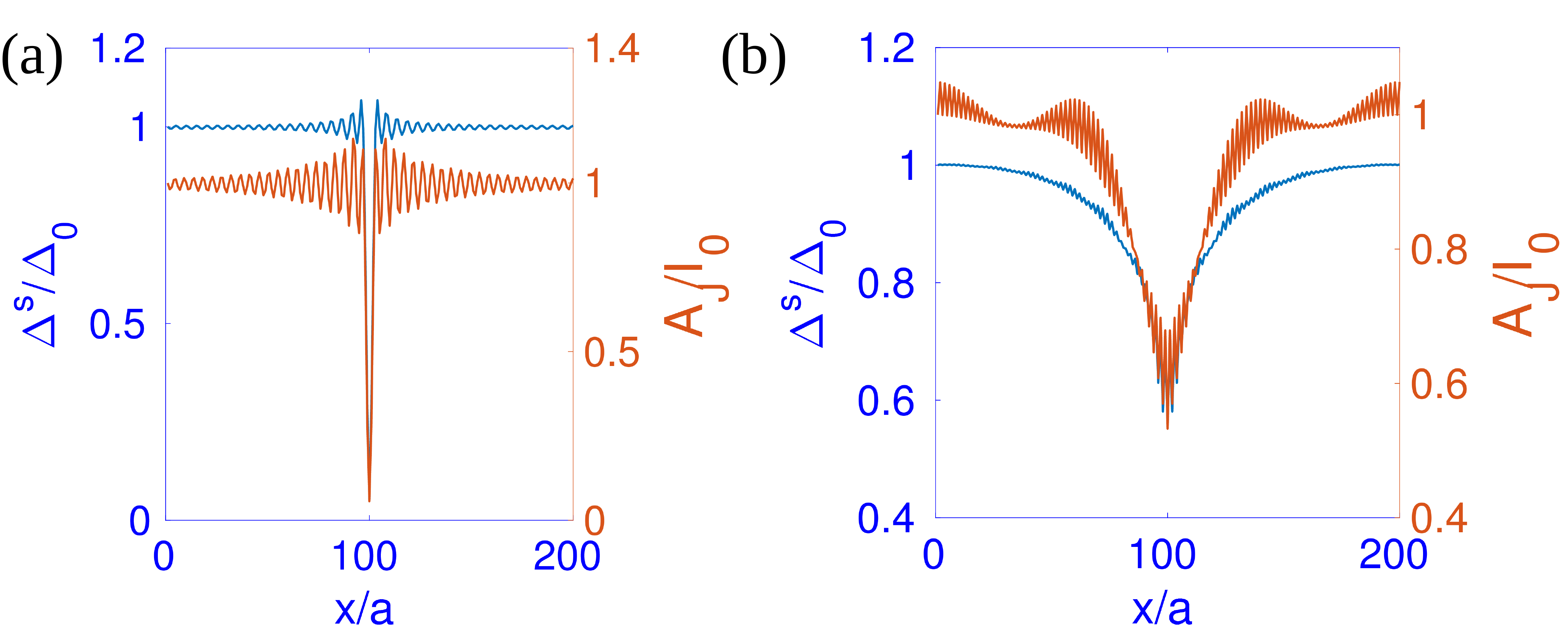}
\caption{Spatial form of the AC-Josephson-current amplitude $A_J$, normalized to its value $I_0$ for the clean case and of the order parameter (blue) for an $s$-wave superconductor with (a) a non-magnetic impurity at site $i_0=100a$, for $\mu_s = -t$, coupling constant $V^s = 0.5$, $U_{\text{imp}}=1t$, $\xi_s \sim 14a$, $\Delta^{\text{tip}}=3\Delta_0^s$, $eV=0.005t$ and (b) a magnetic impurity at site $i_0=100a$, for $\mu_s = 0$, $V^s = 0.5$, $U_{\text{imp}}=0.5t$, $\xi_s \sim 54a$, $\Delta^{\text{tip}}=3\Delta_0^s$, $eV=0.005t$. }
\label{fig:IJeV}
\end{figure}

The result of the numerical evaluation of the general time- and voltage-dependent expression for the Josephson current given in \equref{eq:IJ} for magnetic and non-magnetic defects is illustrated in \figref{fig:IJeV} and compared to the respective local structures of the $s$-wave order parameter. Here, we show the amplitude $A_J$ of the current oscillations, $I_J(t)=A_J\cdot \sin(2eVt - \text{const.})$.
Both quantities, the local structure of the order parameter $\Delta^{s}$ and the Josephson current amplitude $A_J$, have been normalized to their respective values in the absence of defects ($\Delta_0$ and $I_0$, respectively). The Friedel oscillations induced in the local structure of the $s$-wave order parameter by a non-magnetic defect are well mapped by the Josephson current, similar to the zero voltage case considered above.
\begin{figure}[tb]
\includegraphics[width=\linewidth]{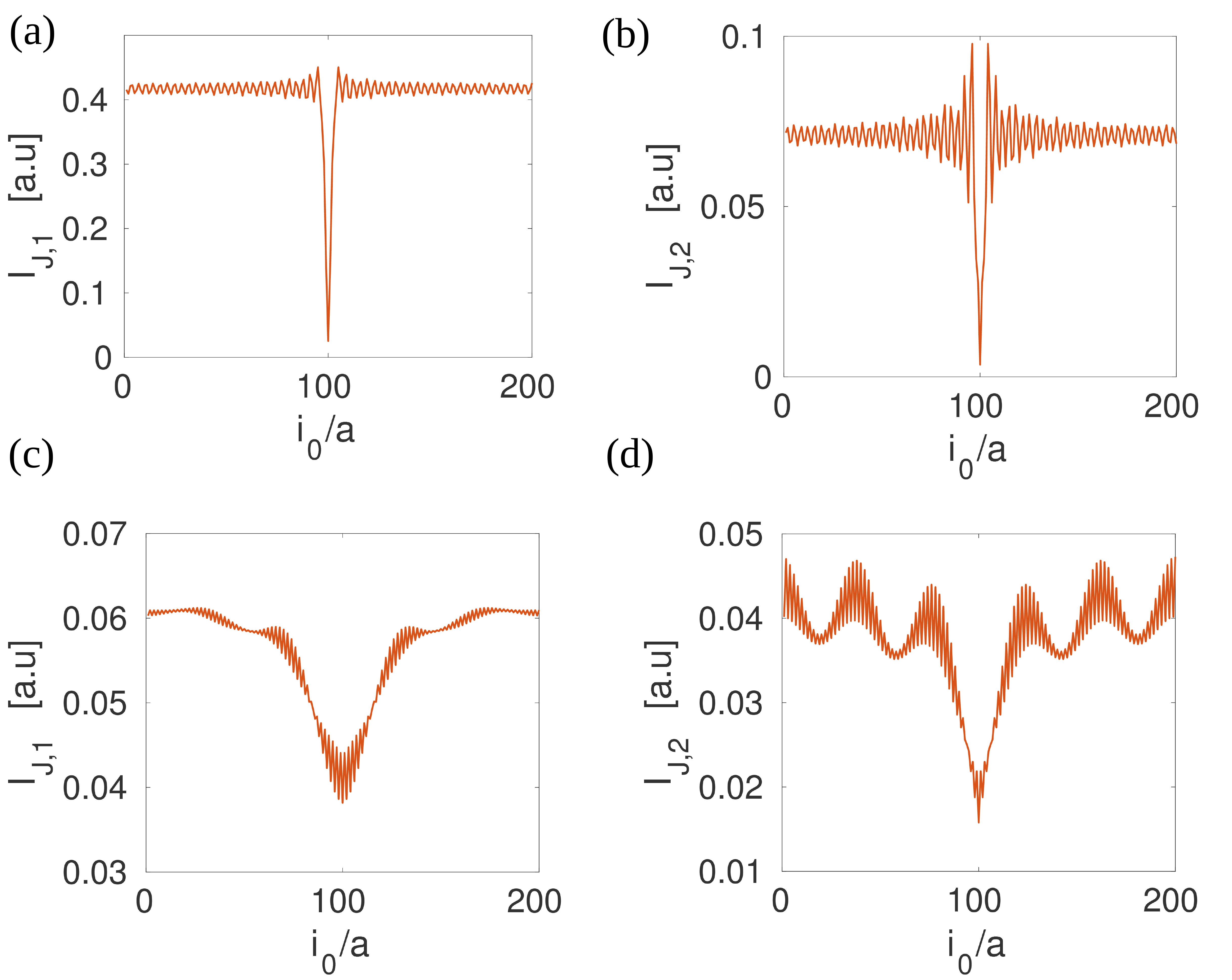}
\caption{(a) Prefactor of the sine-term ($I_{J,1}$) and (b) of the cosine-term ($I_{J,2}$) contribution to the tunneling Josephson current in \equref{eq:IJ1} for an $s$-wave superconducting system with a single non-magnetic defect at $i_0=100a$. Part (c) and (d) show the analogous plots for a magnetic defect at $i_0 = 100a$. System parameters as in \figref{fig:IJeV}.}
\label{fig:IJContributions}
\end{figure}
We find distinct beat features in the Josephson current for the magnetic defect case, which are strongly enhanced compared to zero voltage. The separate contributions $I_{J,1}$ and $I_{J,2}$ to the Josephson current at $eV\neq 0$ are depicted in \figref{fig:IJContributions}. We see that the beat features arise predominantly from the cosine contribution $I_{J,2}$ while being much less pronounced in $I_{J,1}$. 
\subsection{Josephson current for degenerate ground states}
As the Green's function formalism used above is not able to capture the JSTM signature of the MBSs, due to the induced ground state degeneracy, we calculate the current-phase relation directly from\cite{Huang}
\begin{equation}
I_J(\Delta\phi) =  \sum_n \partial E_n(\Delta\phi)/\partial \Delta\phi,
\end{equation}
i.e., by partial derivation of the negative quasi-particle energy spectrum $E_n$ of the JSTM setup with respect to the relative phase $\Delta\phi$ (we set $e=1$ for the electron charge).  As before, we assume a homogeneous $s$-wave superconductor for the tip and take a weak tunneling element of $T_0=0.01t$.

\begin{figure*}[tb]
\includegraphics[width=\linewidth]{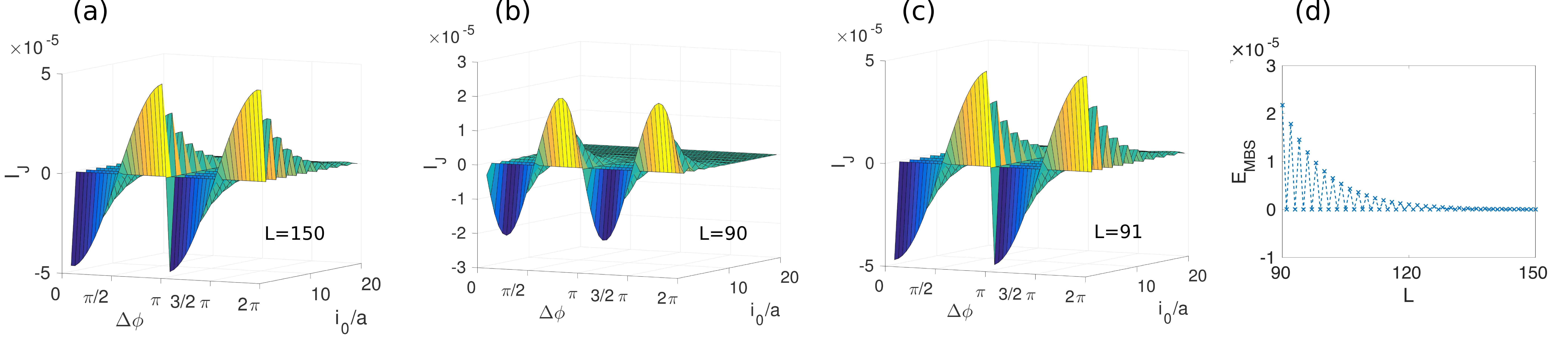}
\caption{Phase $\Delta\phi$ and site $i_0$ dependence of Josephson current for homogeneous $p$-wave superconductor ($\Delta^t=0.1t$) with (a) $L=150$, (b) $L=90$, and (c) $L=91$ sites with $\Delta_{\text{tip}}=0.1t$. The energy of MBSs as a function of system size $L$ is shown in (d).}
\label{fig:pevenodd}
\end{figure*}

\subsubsection{Homogeneous $p$-wave}
The current-phase relation $I_J(i_0,\Delta\phi)$ (with $eV=0$) close to the boundary for a spinfull homogeneous $p$-wave chain, i.e. spatially constant order parameter, is depicted in \figref{fig:pevenodd}(a). Here $i_0$ is the position measured from the sample edge. As a consequence of the distinct parity of the superconductivity in the tip ($s$-wave) and in the superconductor of interest ($p$-wave), there is no Josephson current in the bulk (for $|i_0| \gg \xi_t$). However, due to the presence of MBSs at the edge of the topological $p$-wave state there is a non-zero MBS-induced contribution to Josephson tunneling for $|i_0| \ll \xi_t$, which shows the characteristic\cite{TRSMajoranaJosephson} $\pi$-periodic behavior with discontinuity at $\Delta\phi=\pi$.

We find that the phase relation is sensitive to the degree to which the MBSs at the two ends of the system are separated, i.e., it depends on the ratio of coherence length $\xi_t$ and system size $L$. In \figref{fig:pevenodd}(b) and (c) we show the phase relation for two systems with overlapping MBSs with even and odd number of sites, respectively.
For an even number of sites (see \figref{fig:pevenodd}(b)) the Josephson current retains its $\pi$-periodicity, but the maxima are reduced and shifted and the discontinuity is removed. For an odd number of sites (see \figref{fig:pevenodd}(c)) we regain the current-phase relation of the system with well separated MBSs.
This even-odd-effect mirrors the oscillation of the MBS energies for small system sizes $L$, which is illustrated in \figref{fig:pevenodd}(d).

In \figref{fig:pevenodd}, we observe a second even-odd effect not as a function of system size but as a function of $i_0$, which persists in the thermodynamic limit $L\rightarrow \infty$. The oscillation of the Josephson current amplitude with $i_0$ follows the spatial profile of the MBS wavefunction which has large (small) weight at odd (even) sites $i$.

\begin{figure}[b]
\includegraphics[width=\linewidth]{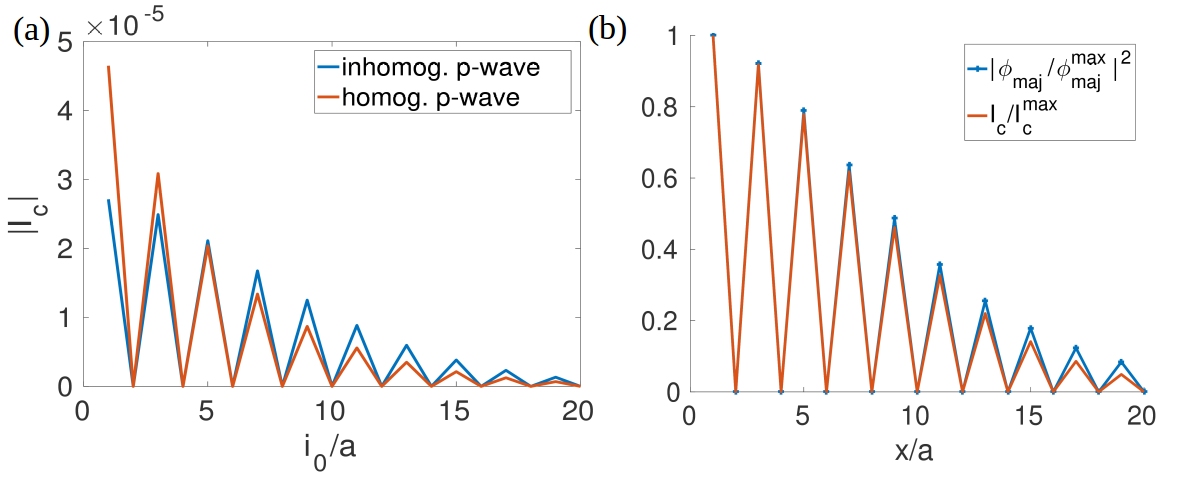}
\caption{Comparison of (a) the critical Josephson current $I_c(i_0)=I_J(i_0,\Delta\phi=0)$ for  spinfull, spatially homogeneous (red) and inhomogeneous (blue) $p$-wave order parameters and of (b) $I_c$ and the modulus squared  $|\phi_{\text{maj}}|^2$ of the MBS wave function for the inhomogeneous $p$-wave state (normalized to their respective maximum values $I_c^{\text{max}}$ and $|\phi^{\text{max}}_{\text{maj}}|^2$).}
\label{fig:Ijcomp}
\end{figure}

\begin{figure}[tb]
\includegraphics[width=\linewidth]{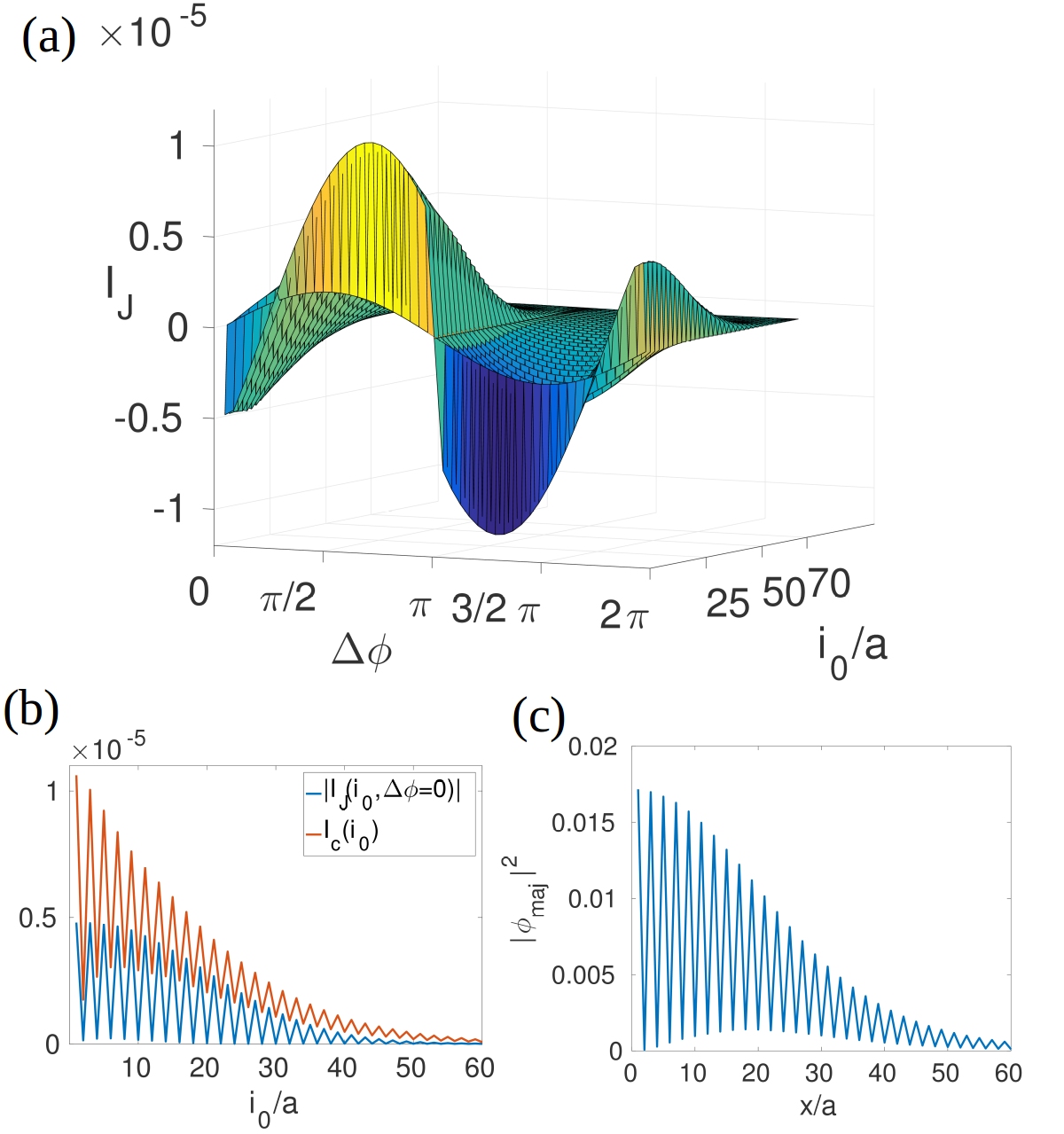}
\caption{(a) Current-phase relation close to the boundary for a superconductor of $L=350$ sites with competing order parameters ($V^t=0.6$, $V^s=0.5$, $\epsilon_c=0.3t$) and bulk dominant $p$-wave ($\Delta^t_0\simeq 0.045$) assuming $\Delta_{\text{tip}}=0.05t$. (b) Spatial profile of Josephson current $|I_J(i_0,\Delta\phi=0)|$ and of critical Josephson current $I_c=I_j(i_0,\Delta\phi\simeq \frac{3}{4}\pi)$ and (c) the modulus squared of the MBS wave function $|\phi_{\text{maj}}|^2$}
\label{fig:spIj350}
\end{figure}

\subsubsection{Self-consistent $p$-wave}
As was discussed in \secref{ImpactOnMBSsWF}, for a spinfull system with inhomogeneous $p$-wave, i.e. spatially varying order parameter, the MBSs are slightly distorted at the system boundaries compared to the homogeneous case.
Comparing the critical Josephson current $I_c(i_0)=I_J(i_0,\Delta\phi=0)$ for these two cases, we find this distortion to be clearly reflected in the Josephson current (see \figref{fig:Ijcomp}(a)), where some of the weight of the Josephson current is shifted towards the bulk.

In \figref{fig:Ijcomp}(b), we can further see that the spatial profile of the critical Josephson current $I_c$ very closely follows the absolute square of the MBS wave function. This clearly illustrates the capability of JSTM not only to detect MBSs but also to map the spatial profile of their wave function.

\begin{figure}[tb]
\includegraphics[width=\linewidth]{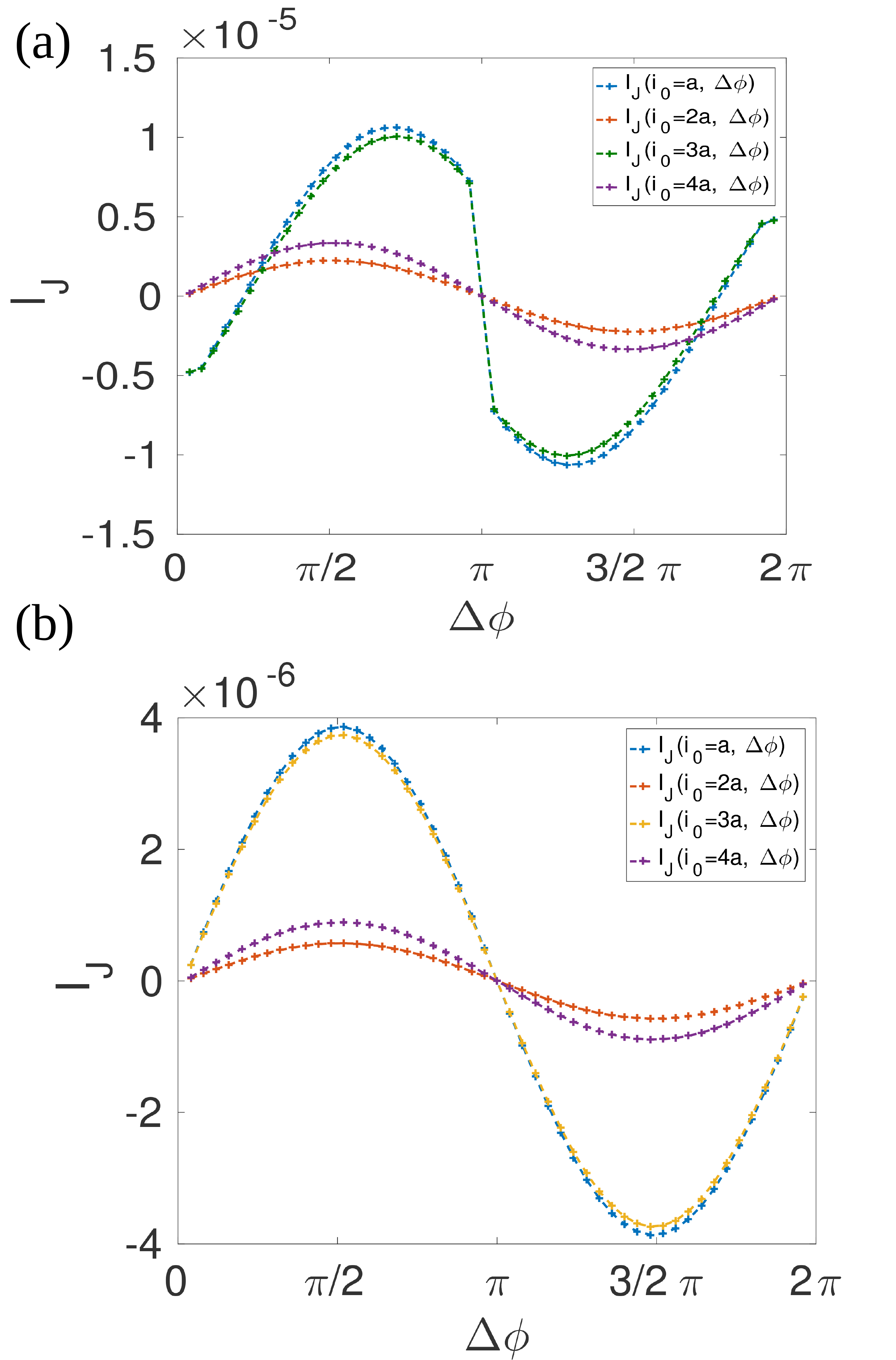}
\caption{Current-phase relation for sites $i_0=a,2a,3a,4a$ for (a) the same parameters as in \figref{fig:spIj350} and (b) for $L=1000$, $V^s=0.35$, $V^t=0.38$ and $\epsilon_c=0.2t$ leading to a much larger coherence length of $\xi_t \approx 165a$. The alternating value of the amplitude of the $\sin(\Delta \phi)$ dependence as a function of $i_0$ in (b) is due to the Friedel-like oscillations of the $s$-wave order parameter.}
\label{fig:Ijsite}
\end{figure}

\subsubsection{Competing order parameters}
Finally, the current-phase relation for a superconductor with competing order parameters with dominant $p$-wave in the bulk and an $s$-wave state emerging at the boundary is shown in \figref{fig:spIj350}. It exhibits a non-zero current contribution for $\Delta\phi=0,2\pi$ and a discontinuity at $\Delta\phi=\pi$. The parameters have been set to yield a bulk $p$-wave gap similar to that of the homogeneous $p$-wave state analyzed above ($\Delta^t_0\simeq 0.1t$). The overall behavior resembles a superposition of the pure $p$-wave contribution, as depicted above, and the well established $\sin(\Delta\phi)$ dependence of a conventional Josephson junction of $s$-wave superconductors. Furthermore, the current decreases towards the bulk, where the $s$-wave component vanishes and the extension of the MBSs ends.

In \figref{fig:spIj350}(b) we show the local Josephson current for $\Delta\phi=0$, which arises solely due to the MBSs, and the critical current $I_c$. The overall form of both quantities is in good agreement with the modulus squared of the MBS wave function in \figref{fig:spIj350}(c). Note that, in contrast to the system examined in \secref{ImpactOnMBSsWF}, the shift of the MBSs away from the edge of the system is not as clear, due to the shorter coherence length (only $\xi_t=24a$ instead of $\xi_t =210a$ in \figref{fig:MBSComp}).

A closer examination of the site- and phase-dependent Josephson current in \figref{fig:spIj350} reveals an alternating pattern between an $s$-wave-dominated profile ($\propto \sin(\Delta\phi)$) of the Josephson current and an admixture of a MBS-induced and of an $s$-wave-like current-phase relation (see \figref{fig:Ijsite}(a)). This alternation is in accordance with the weights of the MBS wave functions (see \figref{fig:spIj350}(c)) being greatest on odd sites and minimal on even sites. For even $i_0$ with weak weights of the MBSs, primarily an $s$-wave signature is seen as the MBS contribution is very small.
For odd $i_0$, the admixture of $\propto \sin(\Delta\phi)$ and of MBS-mediated tunneling leads to a current-phase relation with discontinuous jumps, due to the MBSs, but $2\pi$ periodicity, instead of the expected period of $\pi$, resulting from the additional $s$-wave component.

Note that the contribution of the MBSs to the Josephson current very close to the boundary, $i_0 \ll \xi_t$, is much smaller for systems with larger coherence lengths $\xi_t$. As can be seen in \figref{fig:Ijsite}(b), where $\xi_t=165a$, the Josephson current only shows the conventional $\sin(\Delta\phi)$ form close to the edge. This is due to shift of the MBS wave function away from the edge of the system (see \figref{fig:MBSComp}(a) and (b)) leading to a strong suppression of the MBS contribution to the Josephson current for $i_0 \ll \xi_t$.  While the MBSs can still be resolved in a JSTM setup, a regular Josephson junction between the edge of the topological superconductor and a conventional $s$-wave state, see \figref{BasicSetup}, can hardly detect any signatures of the MBSs.

\section{Conclusions}
In this paper, we have studied the modulations of the superconducting order parameter in the vicinity of edges, magnetic and non-magnetic impurities by self-consistently solving the gap equations for multiple pairing interactions. We have shown that the presence or absence of Friedel oscillations of the superconducting order parameter crucially depends on the pairing symmetry and the nature of the inhomogeneity. Hence, they can be used to discern, on the one hand, between different defect types and, on the other hand, between even parity singlet and odd parity triplet pairing.
This distinction is especially prominent in low-dimensional systems. The $s$-wave order parameter exhibits pronounced Friedel oscillations in the vicinity of non-magnetic inhomogeneities, while the $p$-wave order parameter lacks such distinct features.
We note that a different alternative route to access the character of the order parameter in noncentrosymmetric superconductors, which is based on Shiba states, has been recently proposed in \refcite{PSimon}. 

Furthermore, we have discussed the competition of order parameters with different pairing symmetries in the presence of inhomogeneities. We have shown that this competition can lead to the emergence of a topologically trivial region close to the boundary of a system that is topologically nontrivial in its bulk. Due to this competition, the MBSs are shifted towards the transition point separating these two regions, as opposed to the clear localization at the system edge for a topologically nontrivial system with homogeneous order parameter.

As a means to probe and resolve the Friedel oscillations induced by inhomogeneities, as well as the spatial structure of competing $s$-wave superconductivity and the MBSs, we have discussed JSTM. The spatially resolved Josephson current provides a precise tool for the mapping of the spatial structure of the $s$-wave order parameter as well as of the MBS signatures in systems with competing order parameters. MBSs contribute significantly to the current-phase relation of the Josephson current, modifying it beyond the well established sinusoidal dependence of pure $s$-wave superconductivity, thus revealing the presence of MBSs and their spatial profile. 

Taken together, the spatially resolved Josephson current can be utilized to distinguish between different order parameter symmetries by means of their response to inhomogeneities, discern between magnetic and non-magnetic defects, identify the competition of order parameters, and carry clear signatures of MBSs.
\acknowledgments
MS acknowledges support from the German National Academy of Sciences Leopoldina through grant LPDS 2016-12. This work was supported by the Virtual Materials Design initiative within the Helmholtz program Science and Technology of Nanosystems at the Karlsruhe Institute of Technology.
\bibliographystyle{apsrev4-1}
\bibliography{BdGInhom}
\end{document}